\RequirePackage{fix-cm}
\documentclass[smallcondensed]{svjour3}
\smartqed
\usepackage[utf8]{inputenc}
\usepackage[T1]{fontenc}
\usepackage{times,graphicx,amsmath,amssymb,amsfonts,enumerate,cite,float,marvosym,geometry,needspace}
\usepackage{mathtools}
\usepackage{xcolor}
\usepackage{caption,booktabs,multirow,longtable,chngcntr}
\usepackage[colorlinks,bookmarksopen,bookmarksnumbered,citecolor=blue,linkcolor=blue,urlcolor=blue]{hyperref}
\graphicspath{{./}{article-en/}}
\allowdisplaybreaks[2]
\counterwithout{figure}{section}
\counterwithout{equation}{section}
\counterwithout{table}{section}
\renewcommand{\theequation}{\arabic{equation}}
\renewcommand{\thefigure}{\arabic{figure}}
\renewcommand{\thetable}{\arabic{table}}

\begin{document}
\begin{sloppypar}

\title{Rogue waves in the dam-break problem for the Kundu--Eckhaus equation}
\titlerunning{Rogue waves in the dam-break problem for the Kundu--Eckhaus equation}
\author{Jiaming Sun \and Zhen Wang \and Tianwei Qiu}
\authorrunning{J. Sun et al.}
\institute{J. Sun \and T. Qiu \at School of Mathematical Sciences, Beihang University, Beijing 100191, China\\
\and Z. Wang \Letter \at School of Mathematical Sciences, Beihang University, Beijing 100191, China\\\email{wangzmath@163.com}}
\date{}
\setlength{\headerboxheight}{0pt}
\renewcommand{\makeheadbox}{}
\maketitle

\begin{abstract}

We study rogue wave formation in the dam-break problem for the Kundu--Eckhaus equation with deterministic and randomly perturbed box initial data. For the unperturbed real box, the gauge transformation and Galilean invariance relate the intensity evolution to that of the focusing nonlinear Schr\"odinger equation. The Kundu--Eckhaus parameter determines the phase, the real shift of the discrete spectrum and the drift velocity. We compare uniform, Gaussian and Laplace perturbations with equal variance and the same correlation scale in the underlying Gaussian field. At the larger perturbation strengths considered, the Laplace perturbations yield larger mean maximum intensities and higher exceedance probabilities for the normalized intensity at the final simulation time, but smaller amplification relative to the initial maximum intensity.

\end{abstract}

\section{Introduction}\label{s1}

Rogue waves are transient, localized waves whose peak amplitudes greatly exceed those of the surrounding wave field. Early accounts of unusually large ocean waves include Draper's discussion of freak waves \cite{draper1966freak}. Similar events have been observed in nonlinear optical systems and other dispersive media \cite{AKHMEDIEV20092137,kharif2003physical,onorato2013rogue,solli2007optical,dudley2014instabilities}. Integrable turbulence concerns the statistical evolution of random wave fields governed by integrable systems \cite{zakharov2009turbulence,agafontsev2015integrable}. The nonlinear Schr\"odinger (NLS) equation describes slowly varying wave packets in weakly nonlinear dispersive media \cite{benney1967propagation}. Its coherent structures and statistical properties have been studied in connection with rogue wave formation \cite{akhmediev2009rogue,biondini2015integrable}.

Modulational instability (MI) is a mechanism for rogue wave formation in the focusing NLS equation: small perturbations of a finite-amplitude background grow exponentially \cite{akhmediev2009rogue,biondini2015integrable,gelash2014superregular,biondini2016universal,baronio2015baseband}. For box initial data, the two discontinuities generate oscillatory regions that propagate inward and interact near the center. This dam-break problem has been studied using the inverse scattering transform (IST), Riemann--Hilbert method and Whitham modulation theory \cite{jenkins2014semiclassical,tovbis2016semiclassical,el2016dam}. El et al.~\cite{el2016dam} showed that the interaction produces a modulated breather lattice with localized structures related to rogue waves. Studies of randomly perturbed integrable wave fields have examined extreme events and the resulting wave statistics \cite{agafontsev2015integrable,randoux2014intermittency,walczak2015optical,suret2016single,congy2024statistics}.

The Kundu--Eckhaus (KE) equation is a gauge-equivalent extension of the NLS equation \cite{kundu1984landau,calogero1987nonlinear,kundu2006hierarchy} and takes the form
\begin{equation}
iu_t+u_{xx}+2|u|^2u+4\beta^2|u|^4u-4i\beta(|u|^2)_xu=0,\label{e1}
\end{equation}
where $u(x,t)$ is the complex wave field and $\beta$ is a real parameter. The parameter $\beta$ controls the additional quintic and derivative nonlinearities. Its effects on the phase and spatial evolution are examined below for box initial data. Previous studies of the KE equation include rogue waves \cite{wang2014higher,qiu2015darboux}, solitons \cite{zhu2018riemann,wen2020riemann,xie2018soliton}, modulational instability \cite{parasuraman2021modulational}, chaotic wave fields \cite{bayindir2016chaotic,bayindir2016spectra}, and other analytical and numerical studies \cite{cui2023numerical,li2024integrable,deng2026symmetry}.

Local periodization has been used to characterize individual extreme events through their Floquet spectra \cite{islas2005predicting,schober2022nonlinear,biondini2015integrable,randoux2016inverse}. For the KE equation, Li et al.~\cite{li2024integrable} studied random perturbations of a plane-wave background. However, the dam-break problem with a finite box initial data has received less attention, especially in the presence of random perturbations. In this paper, we compare the effects of different perturbation distributions in the dam-break problem. We use the KE--NLS gauge transformation and Galilean invariance to relate the intensity evolution and scattering data for the unperturbed real box. Numerical simulations test the resulting spectral shift and spatial drift. We then compute the local Floquet spectra of selected rogue waves and compare ensembles generated by uniform, Gaussian and Laplace perturbations. The comparison uses maximum intensities, amplification relative to the initial maximum, intensity statistics and displacement of the discrete eigenvalues.

The remainder of the paper is organized as follows. Section~\ref{s2} derives the KE--NLS correspondence, the MI dispersion relation and the scattering data for the box potential. Section~\ref{s3} examines deterministic dam-break dynamics and the local Floquet spectra of selected rogue waves. Section~\ref{s4} presents the ensemble statistics for randomly perturbed initial data. Section~\ref{s5} discusses the results.

\section{Gauge correspondence and spectral properties}\label{s2}

In this section, we derive the MI dispersion relation for KE plane waves and then relate the KE and NLS box problems through gauge and Galilean transformations. We also obtain the scattering data and discrete spectrum for the KE box potential.

\subsection{Modulational instability of plane waves}

A plane-wave solution of the KE equation is
\[
u_0(x,t)=A e^{i\Omega_0 t},\qquad\Omega_0=2A^2+4\beta^2A^4.
\]
where $A$ is the amplitude and $\Omega_0$ is the frequency. We perturb this solution as
\[
u(x,t)=\left[A+\eta_r(x,t)+i\eta_i(x,t)\right]e^{i\Omega_0 t},
\]
where $\eta_r$ and $\eta_i$ denote the real and imaginary parts of the perturbation, respectively, with $|\eta_r|,|\eta_i|\ll A$. Substituting this expression into the KE equation and retaining terms linear in $\eta_r$ and $\eta_i$ gives
\[
\eta_{r,xx}-\eta_{i,t}+4A^2\left(1+4\beta^2A^2\right)\eta_r=0,
\]
and
\[
\eta_{i,xx}+\eta_{r,t}-8\beta A^2\eta_{r,x}=0.
\]
We seek normal-mode perturbations of the form $\eta_r,\eta_i\propto e^{i(kx-\omega t)}$, where $k$ and $\omega$ denote the perturbation wavenumber and frequency, respectively. A nontrivial solution exists only if
\[
\det
\begin{pmatrix}
-k^2+4A^2+16\beta^2A^4 & i\omega\\[1mm]
-i(\omega+8\beta A^2k) & -k^2
\end{pmatrix}
=0.
\]
This condition gives the dispersion relation
\begin{equation}
\omega=-4\beta A^2k\pm k\sqrt{k^2-4A^2}.\label{eq:MI_dispersion}
\end{equation}

For $|k|<2A$, the frequency has a nonzero imaginary part and the perturbation grows exponentially. The corresponding growth rate is
\begin{equation}
\gamma(k)=|k|\sqrt{4A^2-k^2}.\label{eq:MI_growth}
\end{equation}
The growth rate in Equation~\eqref{eq:MI_growth} attains its maximum at
\[
|k|=\sqrt{2}A,\qquad\gamma_{\max}=2A^2.
\]
The unstable wavenumber interval and growth rate are independent of $\beta$ and coincide with those for the focusing NLS equation. In Equation~\eqref{eq:MI_dispersion}, $\beta$ contributes the real frequency shift $-4\beta A^2k$, with convective velocity $v_{\mathrm{MI}}=-4\beta A^2$. Thus, $\beta$ changes the transport velocity of the unstable modulation but not its growth rate.

\subsection{The KE--NLS correspondence}

The KE and NLS equations are related through a gauge transformation. Let $q(x,t)$ be a solution of the focusing NLS equation
\begin{equation}
iq_t+q_{xx}+2|q|^2q=0.\label{eq:NLS}
\end{equation}
For the localized wave fields considered here, define the real phase
\begin{equation}
\Theta(x,t)=2\beta\int_{-\infty}^{x}|q(y,t)|^2\,dy.\label{eq:gauge_phase}
\end{equation}
The KE field is then given by
\begin{equation}
u(x,t)=q(x,t)e^{i\Theta(x,t)}.\label{eq:gauge_transform}
\end{equation}

The NLS equation~\eqref{eq:NLS} satisfies the local mass conservation law
\[
\partial_t|q|^2=i\partial_x\left(q^*q_x-qq_x^*\right).
\]
Together with Equation~\eqref{eq:gauge_phase}, this relation gives
\[
\Theta_x=2\beta|q|^2,\qquad\Theta_t=2i\beta\left(q^*q_x-qq_x^*\right).
\]
The field defined by \eqref{eq:gauge_transform} satisfies the KE equation~\eqref{e1}, as follows by substitution using \eqref{eq:NLS} and the phase derivatives above \cite{kundu1984landau,calogero1987nonlinear}. Since the gauge factor has unit modulus, $|u(x,t)|=|q(x,t)|$. The gauge transformation therefore preserves the amplitude and intensity and changes only the phase.

For the KE dam-break problem, we take the box initial data
\begin{equation}
u(x,0)=
\begin{cases}
1, & |x|\leq l,\\
0, & |x|>l,
\end{cases}\label{cz}
\end{equation}
where $l$ is the box half-width. With the lower integration limit in Equation~\eqref{eq:gauge_phase} fixed at $-\infty$, the phase inside the box is
\[
\Theta(x,0)=2\beta(x+l),\qquad |x|\leq l.
\]
Equation~\eqref{cz} therefore corresponds to the NLS initial data
\begin{equation}
q(x,0)=
\begin{cases}
e^{-2i\beta(x+l)}, & |x|\leq l,\\
0, & |x|>l.
\end{cases}\label{eq:NLS_phase_box}
\end{equation}
Thus, the real KE box maps to an NLS box with the same amplitude and a phase.

A Galilean transformation removes this phase. Let $Q(x,t)$ denote the focusing NLS solution generated by
\[
Q(x,0)=
\begin{cases}
1, & |x|\leq l,\\
0, & |x|>l.
\end{cases}
\]
The Galilean invariance of the focusing NLS equation gives
\begin{equation}
q(x,t)=e^{-2i\beta(x+l)-4i\beta^2t}Q(x+4\beta t,t).\label{eq:Galilean_relation}
\end{equation}
Equations~\eqref{eq:gauge_transform} and \eqref{eq:Galilean_relation} then give $|u(x,t)|=|Q(x+4\beta t,t)|$. Thus, the two dam-break solutions have identical intensities up to a spatial translation. The displacement and drift velocity are
\begin{equation}
x_{\mathrm{drift}}(t)=-4\beta t,\qquad v_{\mathrm{drift}}=-4\beta. \label{eq:drift_law}
\end{equation}
After this translation, the intensity evolution coincides with that of the focusing NLS box problem studied by El et al.~\cite{el2016dam}.

\subsection{Lax spectrum and scattering data for the box potential}

\begin{figure}[htbp]
    \centering
    \includegraphics[width=0.68\linewidth]{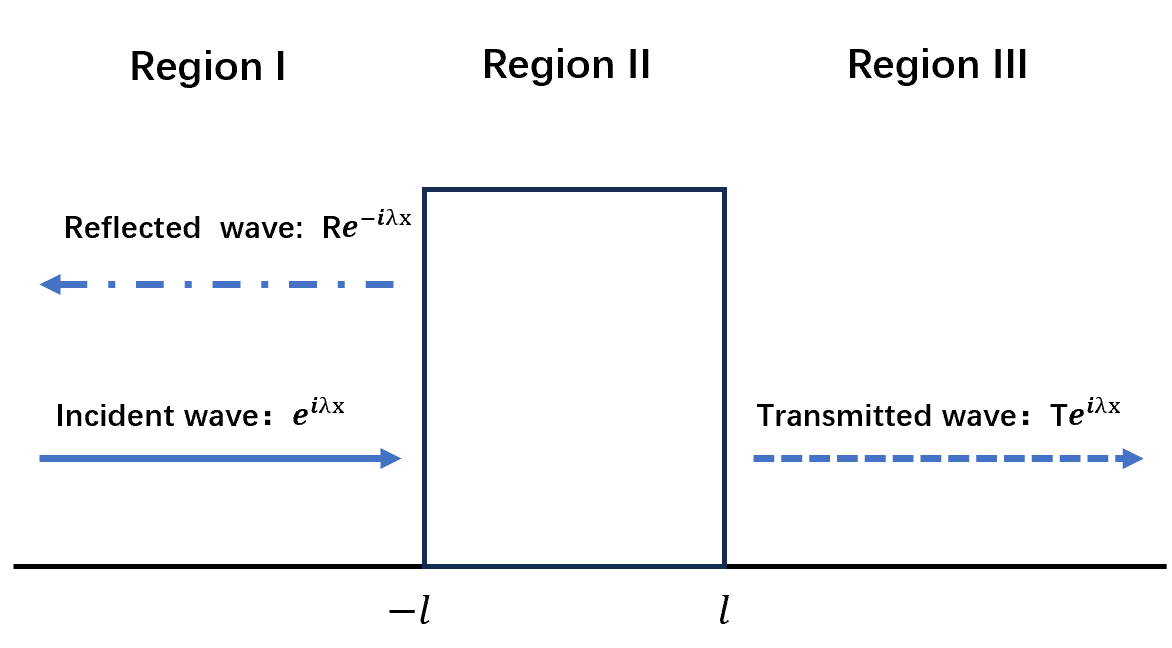}
    \caption{Schematic representation of the direct scattering problem for the box potential. The scattering solutions are matched continuously at$x=-l$ and $x=l$.}
    \label{fig:scattering_schematic}
\end{figure}

The spatial Lax problem for the KE equation is
\begin{equation}
\begin{pmatrix}
-\partial_x+i\beta|u|^2 & u\\
u^* & \partial_x+i\beta|u|^2
\end{pmatrix}
\boldsymbol{\psi}=i\lambda\boldsymbol{\psi},\label{eq:KE_spatial_Lax}
\end{equation}
where $\lambda\in\mathbb{C}$ is the spectral parameter. The induced transformation of the eigenfunction reduces Equation~\eqref{eq:KE_spatial_Lax} to the Zakharov--Shabat problem for the focusing NLS equation. Removing the phase of the NLS box in Equation~\eqref{eq:NLS_phase_box} shifts the spectral parameter to
\begin{equation}
\mu=\lambda-\beta.\label{eq:spectral_shift}
\end{equation}
The spectral transformation and scattering data are derived in \ref{app:scattering_derivation}. We solve the direct scattering problem separately for $x<-l$, $-l\leq x\leq l$ and $x>l$, and match the solutions continuously at $x=\pm l$; see Figure~\ref{fig:scattering_schematic}.

For the shifted spectral parameter $\mu=\lambda-\beta$, the wavenumber inside the box is
\[
\kappa=\sqrt{1+\mu^2}=\sqrt{1+(\lambda-\beta)^2}.
\]
Matching at $x=\pm l$ gives the KE scattering coefficients
\[
a_{\mathrm{KE}}(\lambda)=\frac{\kappa\cos(2\kappa l)-i\mu\sin(2\kappa l)}{\kappa}e^{2i\lambda l},
\]
and
\[
b_{\mathrm{KE}}(\lambda)=-\frac{\sin(2\kappa l)}{\kappa}.
\]
For the real NLS box potential, the scattering coefficients are \cite{jenkins2014semiclassical}
\[
a_{\mathrm{NLS}}(\mu)=\frac{\kappa\cos(2\kappa l)-i\mu\sin(2\kappa l)}{\kappa}e^{2i\mu l},
\]
and
\[
b_{\mathrm{NLS}}(\mu)=-\frac{\sin(2\kappa l)}{\kappa}.
\]
The relation $\lambda=\mu+\beta$ then gives
\begin{equation}
a_{\mathrm{KE}}(\lambda)=e^{2i\beta l}a_{\mathrm{NLS}}(\lambda-\beta),\label{eq:scattering_a_relation}
\end{equation}
and
\[
b_{\mathrm{KE}}(\lambda)=b_{\mathrm{NLS}}(\lambda-\beta).
\]
The KE spectrum is shifted by $\beta$ along the real axis relative to the NLS spectrum for the real box potential. The factor $e^{2i\beta l}$ changes the phase of $a_{\mathrm{KE}}$ but leaves its zeros unchanged.

For real $\lambda$, the scattering data satisfy
\[
|a_{\mathrm{KE}}(\lambda)|^2+|b_{\mathrm{KE}}(\lambda)|^2=1,\qquad\lambda\in\mathbb{R}.
\]
This relation is the reduction condition for the scattering matrix on the real spectral axis. The discrete eigenvalues are determined by $a_{\mathrm{KE}}(\lambda_n)=0$. By Equations~\eqref{eq:spectral_shift} and \eqref{eq:scattering_a_relation}, these zeros are shifted by $\beta$ relative to those for the real NLS box potential. The NLS eigenvalues in the upper half-plane lie on the imaginary $\mu$-axis:
\begin{equation}
\mu_n=i\nu_n,\qquad0<\nu_n<1.\label{eq:NLS_discrete_eigenvalues}
\end{equation}
Equation~\eqref{eq:NLS_discrete_eigenvalues} gives the corresponding KE eigenvalues as
\begin{equation}
\lambda_n=\beta+i\nu_n.\label{eq:KE_discrete_eigenvalues}
\end{equation}

\begin{figure}[htbp]
    \centering
    \includegraphics[width=0.68\linewidth]{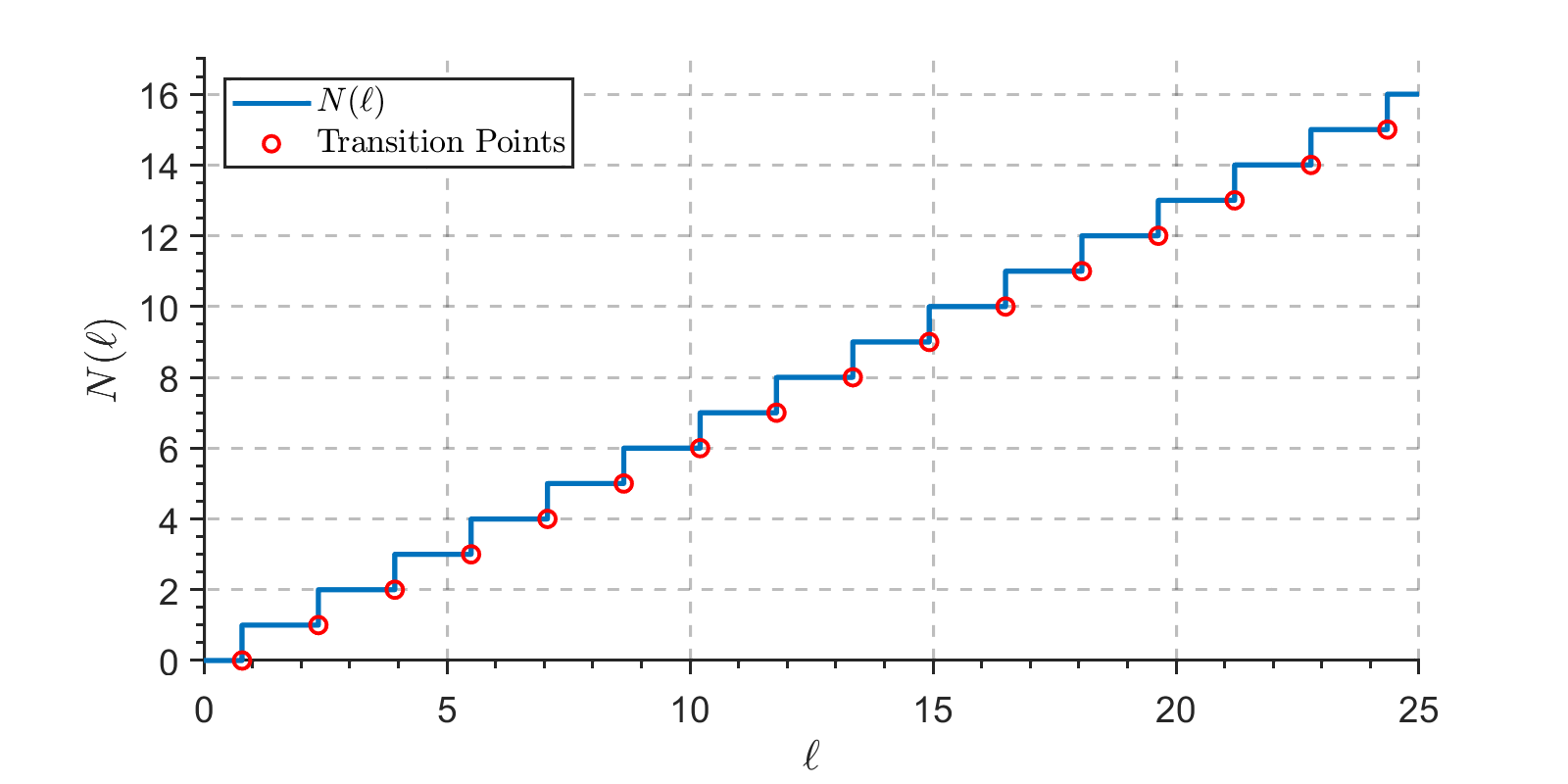}
    \caption{Number of discrete eigenvalues in the upper half-plane as a functionof the half-width $l$ of the box initial data.}
    \label{fig:eigenvalue_count}
\end{figure}

For $\mu=i\nu$, the zero condition becomes
\begin{equation}
\sqrt{1-\nu^2}\cos\left(2l\sqrt{1-\nu^2}\right)+\nu\sin\left(2l\sqrt{1-\nu^2}\right)=0.\label{eq:eigenvalue_condition}
\end{equation}
Equation~\eqref{eq:eigenvalue_condition} is the $\beta$-independent eigenvalue condition for the real NLS box potential. The number of discrete eigenvalues in the upper half-plane is
\begin{equation}
N(l)=
\begin{cases}
0,&0<l\leq\dfrac{\pi}{4},
\\[3mm]
n,&\dfrac{(2n-1)\pi}{4}<l\leq\dfrac{(2n+1)\pi}{4},\qquad n=1,2,\ldots.
\end{cases}\label{eq:eigenvalue_count}
\end{equation}
The counting formula~\eqref{eq:eigenvalue_count} is derived in \ref{app:discrete_count}. Figure~\ref{fig:eigenvalue_count} shows the stepwise increase of $N(l)$ with the box half-width. The transition values of $l$ are independent of $\beta$ and coincide with those for the NLS box problem. Thus, $\beta$ changes neither the number nor the imaginary parts of the discrete eigenvalues; it shifts their real parts uniformly to $\operatorname{Re}\lambda_n=\beta$.

\section{Deterministic dam-break dynamics and local Floquet spectra} \label{s3}

In this section, we examine the dependence of deterministic dam-break dynamics on the box half-width and test the spectral shift and spatial drift derived in Section~\ref{s2}. We then compute local Floquet spectra for selected intense events. 

\begin{figure}[htbp]
\centering
\begin{minipage}{0.24\textwidth}
    \centering
    \includegraphics[width=\linewidth]{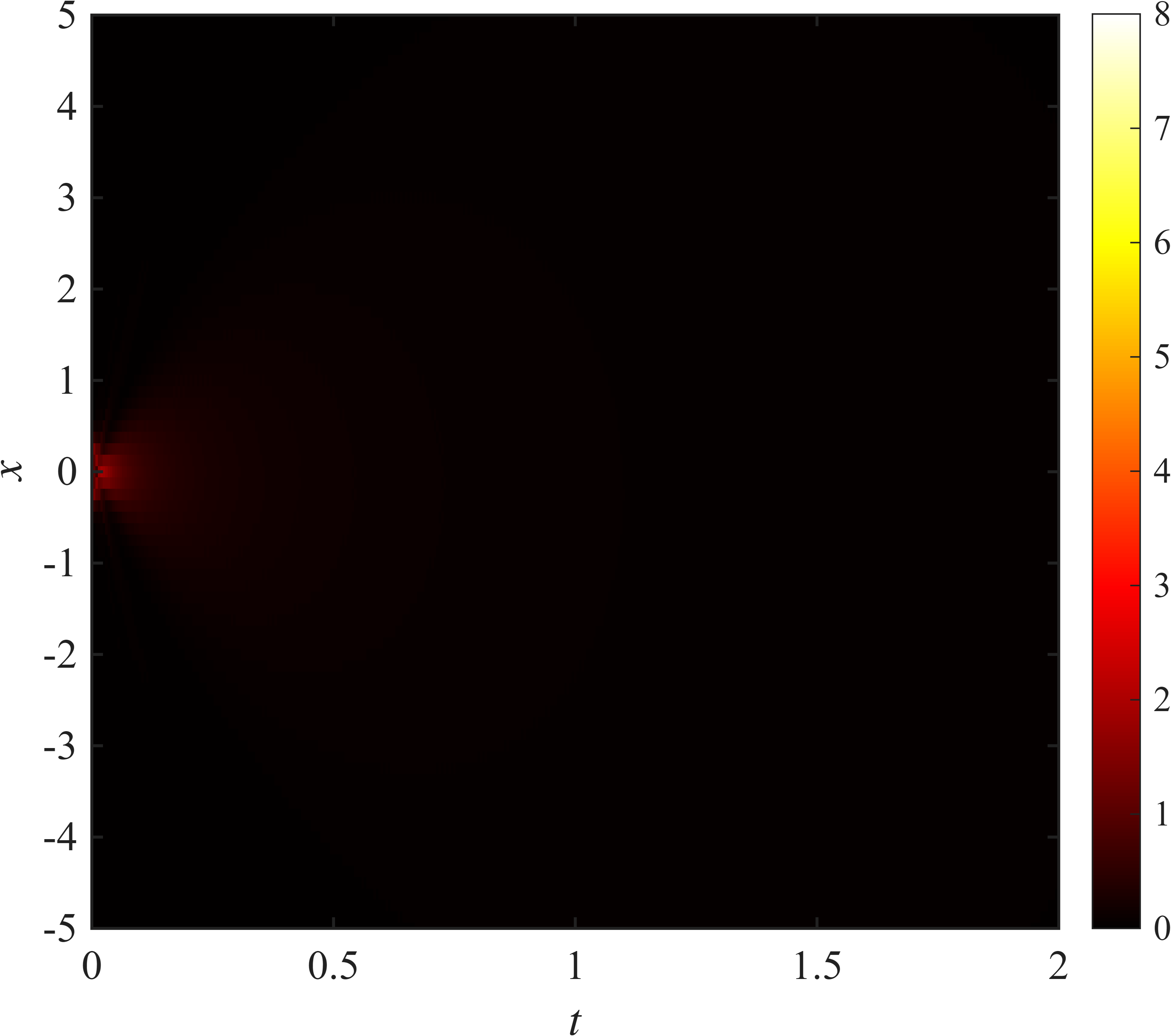}
    \caption*{(a1)}
\end{minipage}
\begin{minipage}{0.24\textwidth}
    \centering
    \includegraphics[width=\linewidth]{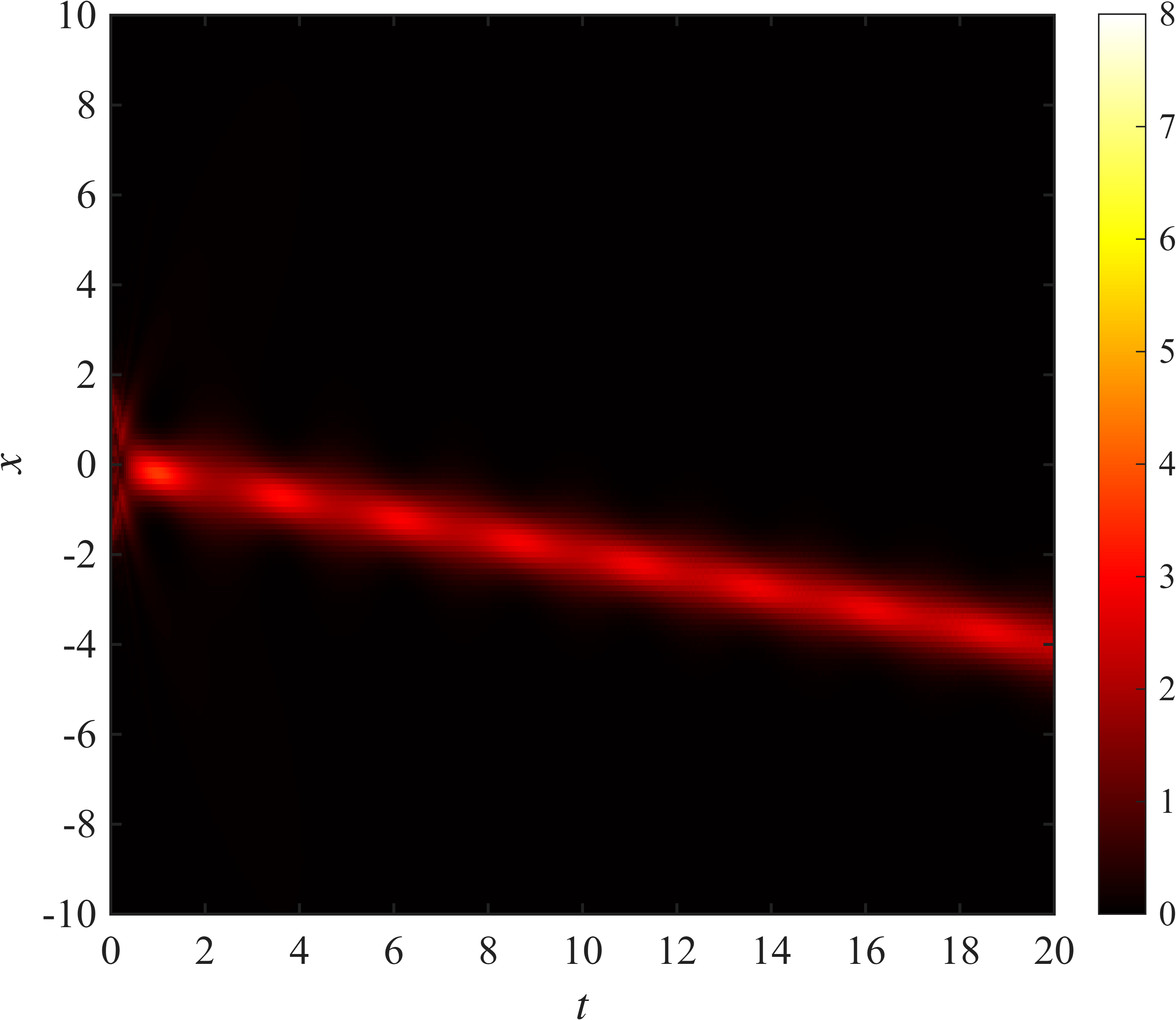}
    \caption*{(b1)}
\end{minipage}
\begin{minipage}{0.24\textwidth}
    \centering
    \includegraphics[width=\linewidth]{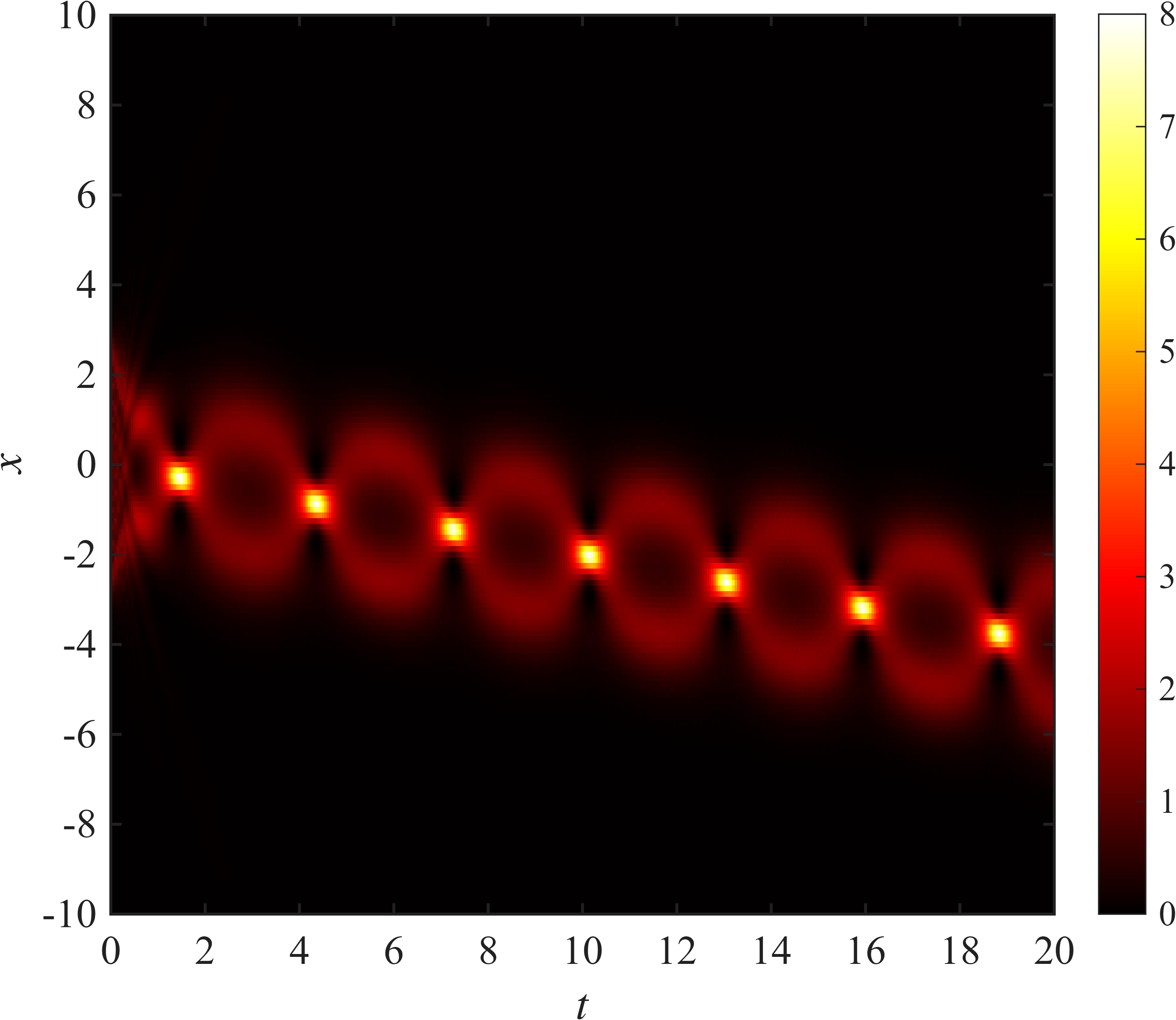}
    \caption*{(c1)}
\end{minipage}
\begin{minipage}{0.24\textwidth}
    \centering
    \includegraphics[width=\linewidth]{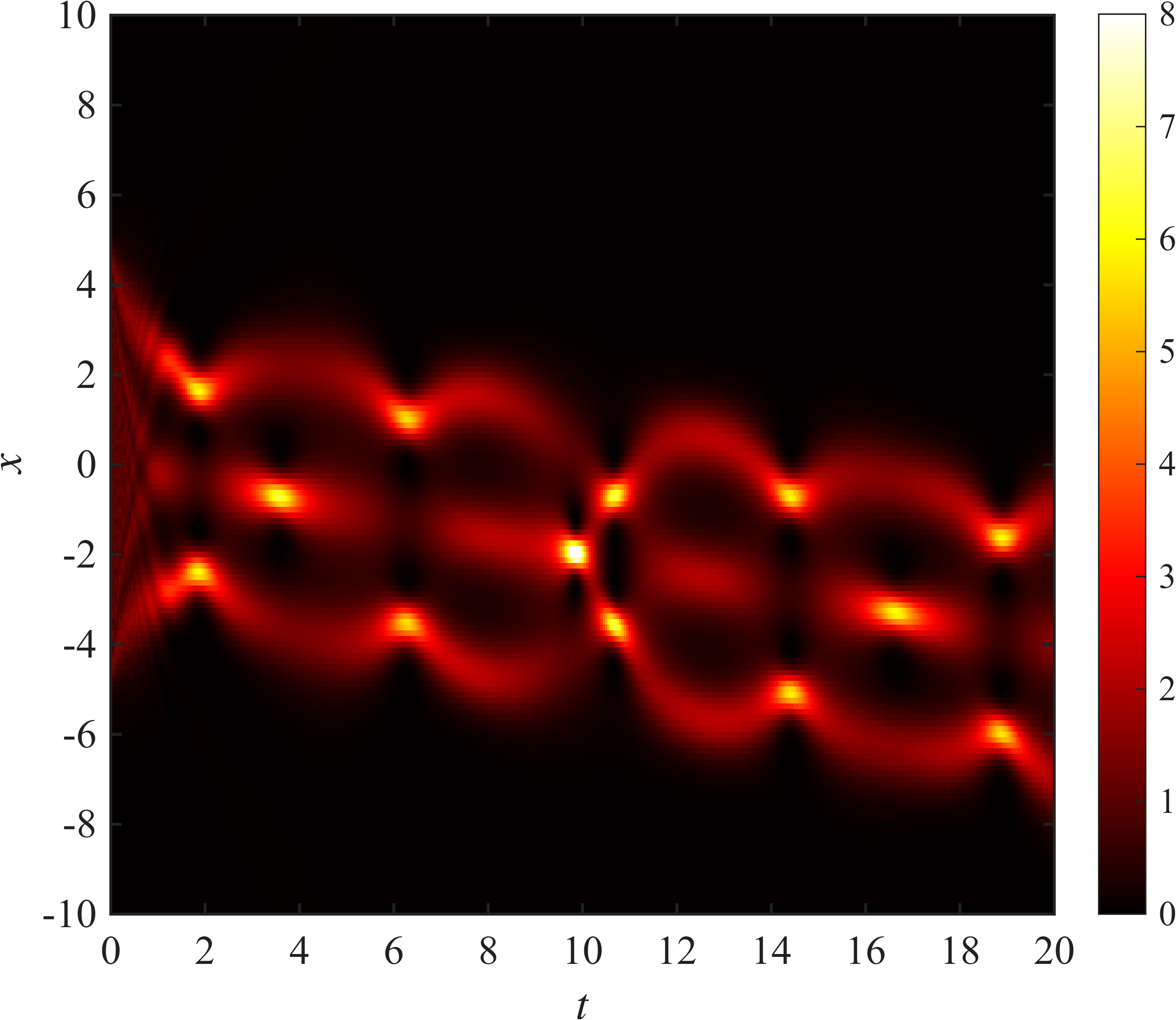}
    \caption*{(d1)}
\end{minipage}

\vspace{0.3cm}

\begin{minipage}{0.24\textwidth}
    \centering
    \includegraphics[width=\linewidth]{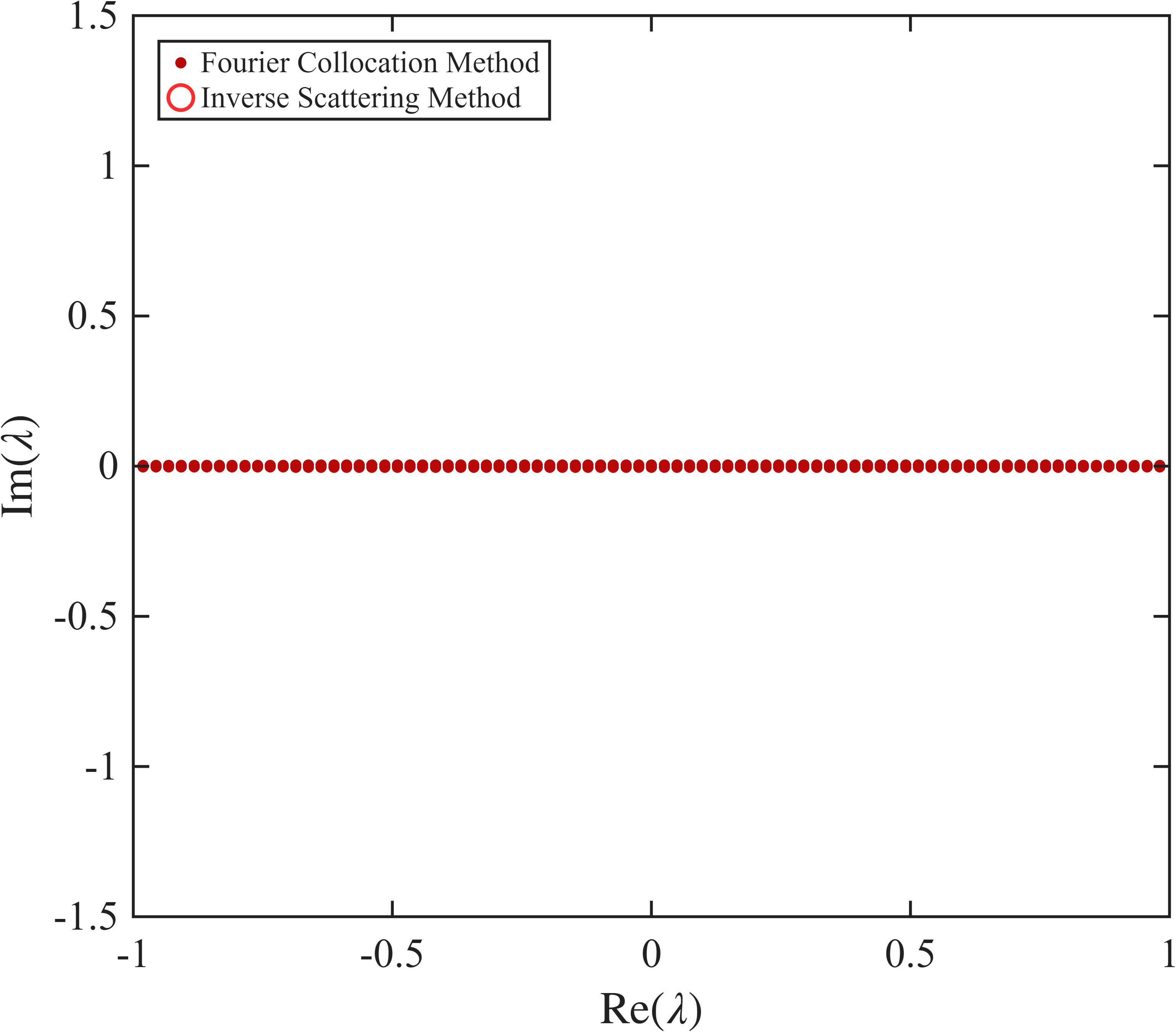}
    \caption*{(a2)}
\end{minipage}
\begin{minipage}{0.24\textwidth}
    \centering
    \includegraphics[width=\linewidth]{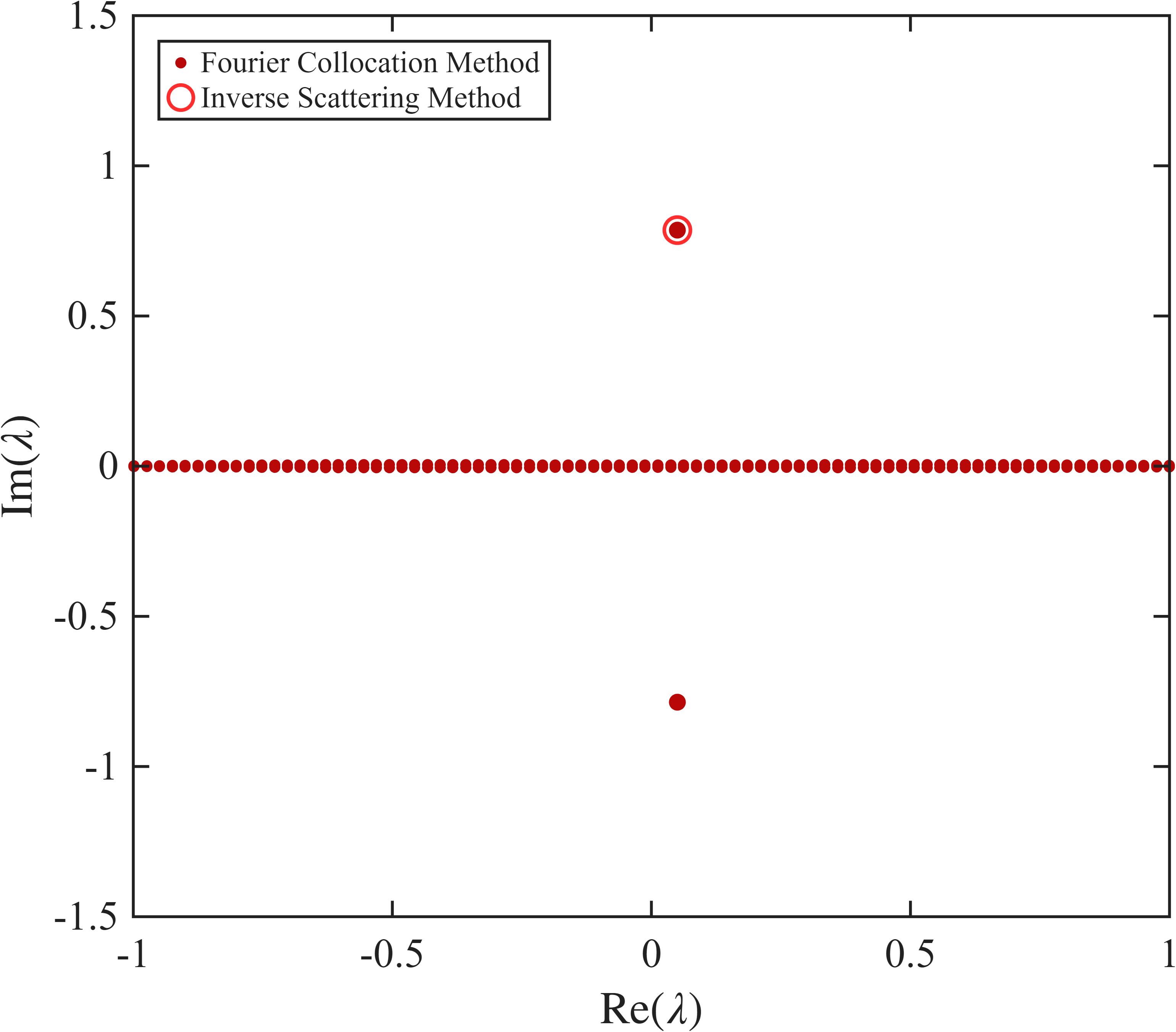}
    \caption*{(b2)}
\end{minipage}
\begin{minipage}{0.24\textwidth}
    \centering
    \includegraphics[width=\linewidth]{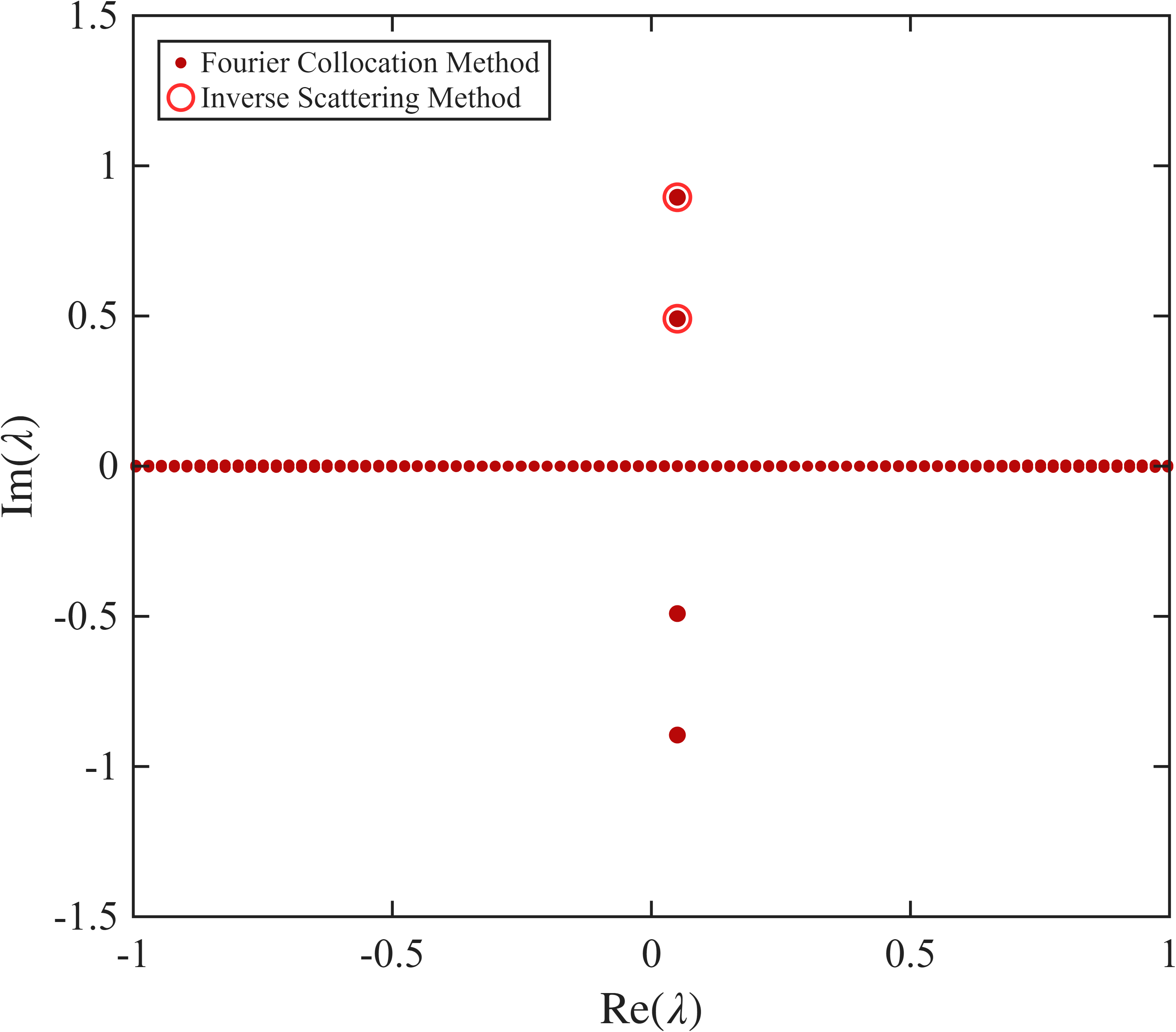}
    \caption*{(c2)}
\end{minipage}
\begin{minipage}{0.24\textwidth}
    \centering
    \includegraphics[width=\linewidth]{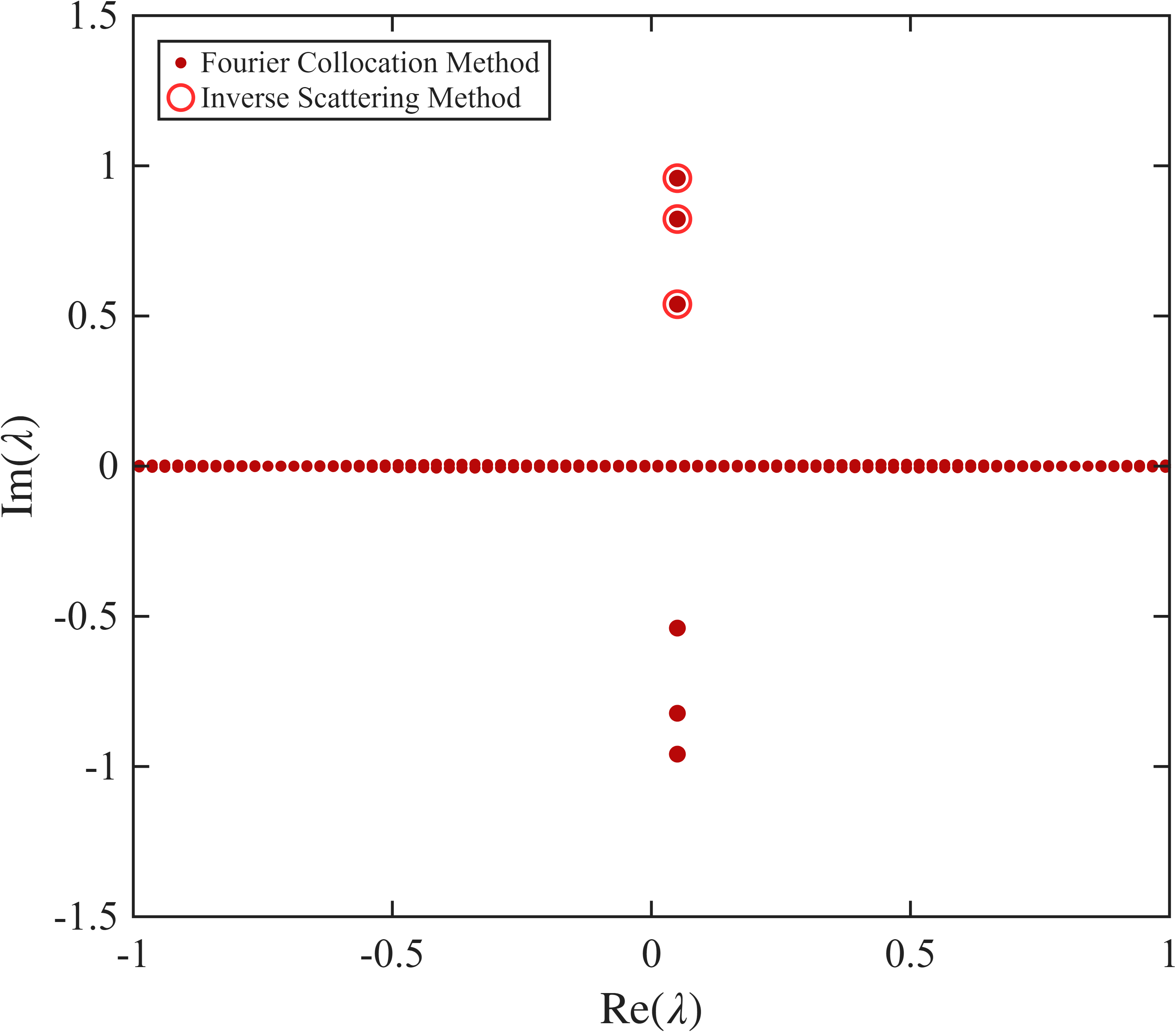}
    \caption*{(d2)}
\end{minipage}

\caption{Space-time evolution and corresponding Lax spectra for different values of the box half-width at $\beta=0.05$. (a1)--(a2) $l=0.4$; (b1)--(b2) $l=2$; (c1)--(c2) $l=3$; (d1)--(d2) $l=5$. The spectra are computed using Fourier collocation with $L=256$ and $N=1024$ and are compared with the discrete eigenvalues obtained from the direct scattering.}
\label{fsp}
\end{figure}

For $\beta=0.05$, we evolve the KE box initial data~\eqref{cz} with $l=0.4$, $2$, $3$ and $5$. Figure~\ref{fsp} shows the resulting space-time evolution and Lax spectra. The computed spectra contain no discrete eigenvalue for $l=0.4$ and one for $l=2$, with additional eigenvalues appearing for $l=3$ and $5$, as predicted in Section~\ref{s2}. The evolution for the wider boxes contains more interacting coherent structures and localized peaks, consistent with the NLS dam-break dynamics reported by El et al.~\cite{el2016dam}.

For $l=5$, we test the discrete spectral shift $\lambda_n=\beta+i\nu_n$, predicted by Equation~\eqref{eq:KE_discrete_eigenvalues}, where $\nu_n$ is independent of $\beta$. We use $\beta=-0.10,-0.05,0,0.05,0.10$. Three discrete eigenvalues are obtained for each value of $\beta$, in agreement with the theoretical prediction $N(5)=3$. Table~\ref{tab:beta_spectral_shift} summarizes the numerical errors. For all tested values of $\beta$, the real parts agree with $\operatorname{Re}\lambda_n=\beta$ to within $1.23\times10^{-6}$, and the imaginary parts vary by less than $10^{-7}$, consistent with the shift $\mu=\lambda-\beta$ derived in Section~\ref{s2}.

\begin{table}[H]
\centering
\caption{Numerical validation of the translation of the discrete eigenvalues along the real axis for $l=5$.}
\label{tab:beta_spectral_shift}
\begin{tabular}{cccc}
\toprule
Mode &
$\operatorname{Im}\lambda_n|_{\beta=0}$ &
$\displaystyle \max_{\beta}|\operatorname{Re}\lambda_n-\beta|$ &
$\displaystyle \max_{\beta}
\left|
\operatorname{Im}\lambda_n(\beta)
-
\operatorname{Im}\lambda_n(0)
\right|$ \\
\midrule
1 & 0.538963 & $1.233\times10^{-6}$ & $9.194\times10^{-8}$ \\
2 & 0.823077 & $6.173\times10^{-7}$ & $7.206\times10^{-9}$ \\
3 & 0.958457 & $1.526\times10^{-7}$ & $6.364\times10^{-9}$ \\
\bottomrule
\end{tabular}
\end{table}

\begin{figure}[htbp]
    \centering
    \begin{minipage}{0.45\textwidth}
        \centering
        \includegraphics[width=\linewidth]{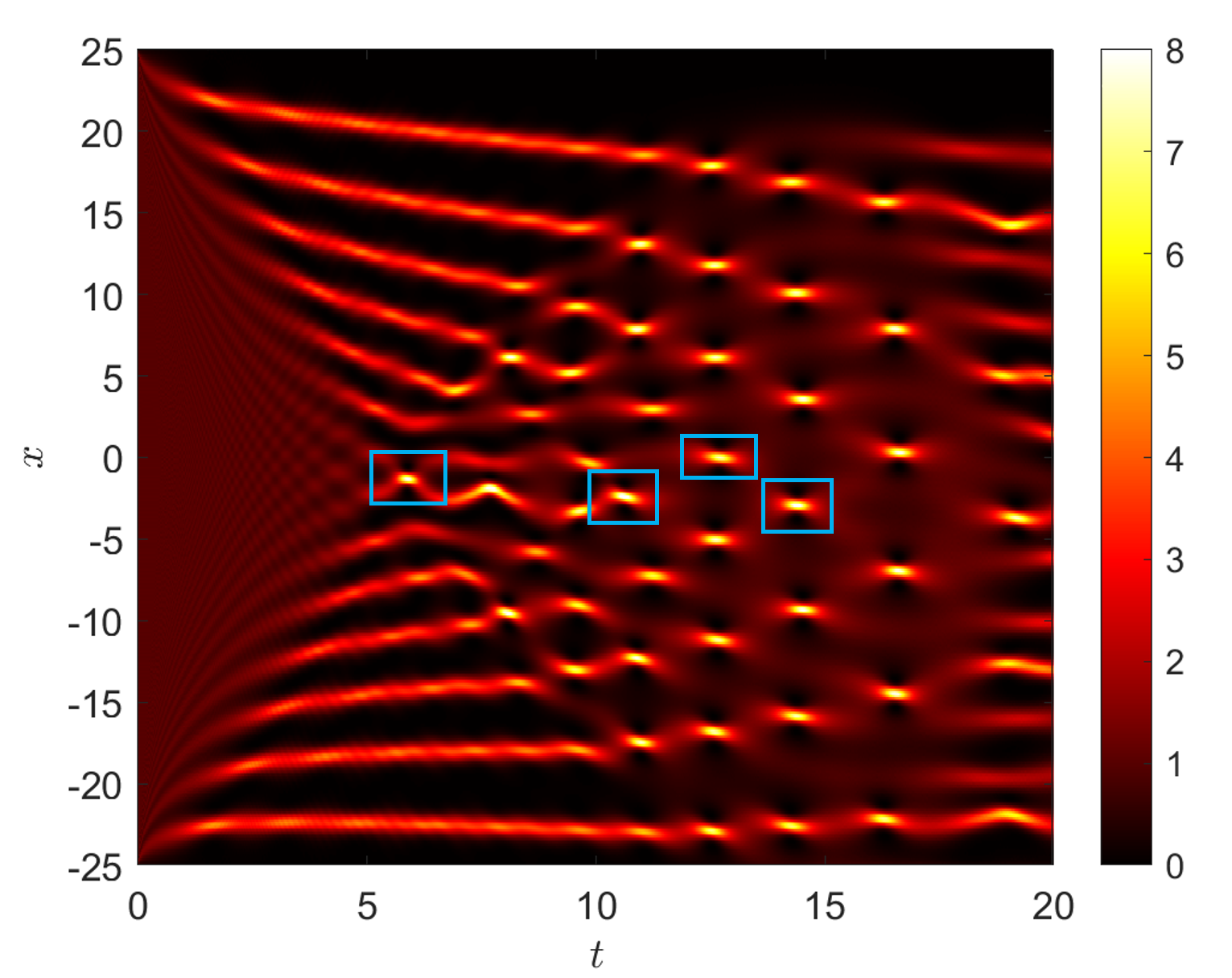}
        \caption*{(a)}
    \end{minipage}
    \begin{minipage}{0.40\textwidth}
        \centering
        \includegraphics[width=\linewidth]{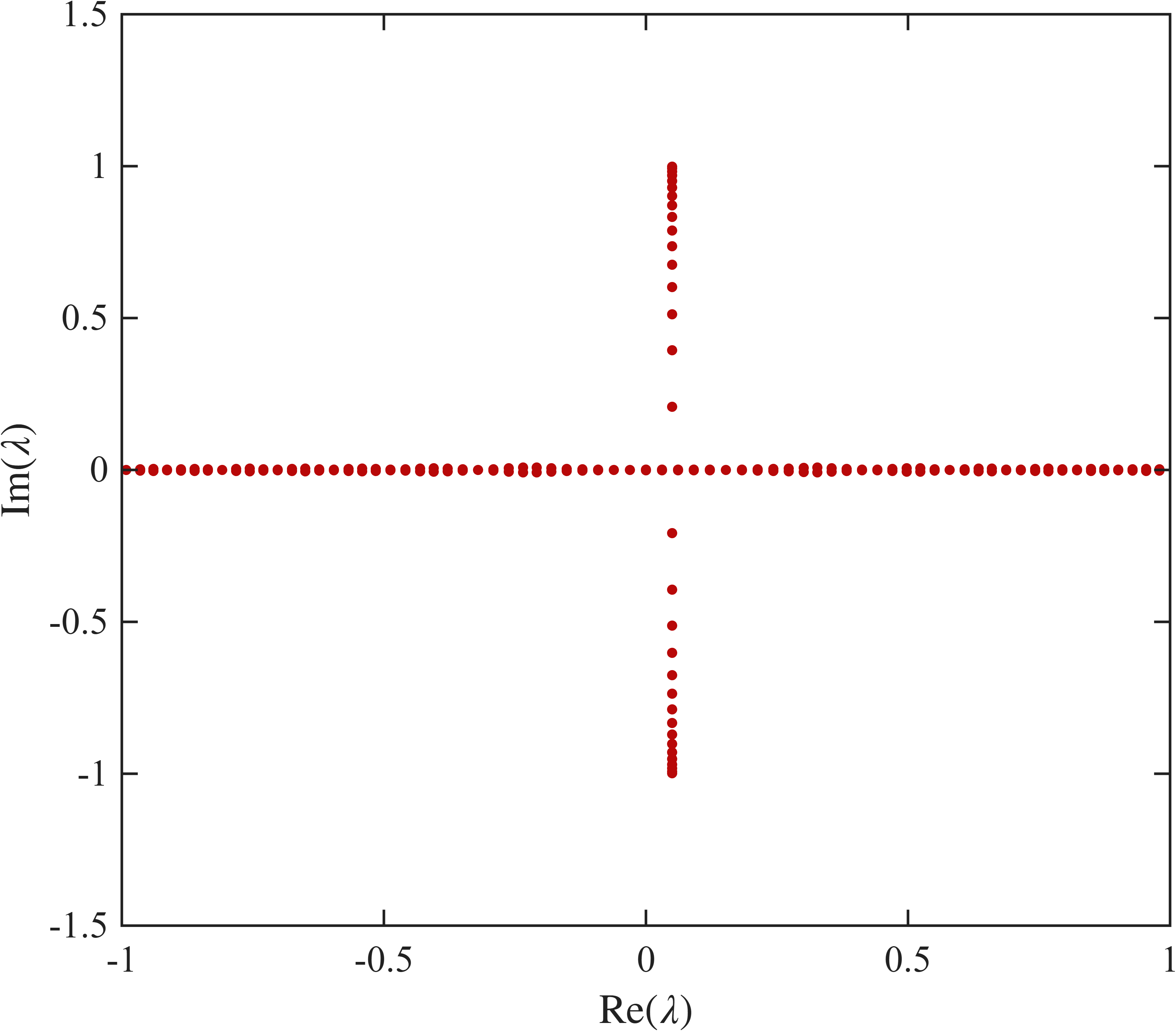}
        \caption*{(b)}
    \end{minipage}
    \caption{Space-time evolution and corresponding Lax spectrum for the dam-break problem with $l=25$ and $\beta=0.05$.}
    \label{f2}
\end{figure}

Figure~\ref{f2}(a) shows the evolution for $l=25$. In the corresponding NLS problem, nonlinear wave trains propagate inward from the box edges and interact near the center \cite{el2016dam}. The KE intensity is obtained by the spatial translation in Equation~\eqref{eq:drift_law}. The corresponding Lax spectrum is shown in Figure~\ref{f2}(b): the discrete eigenvalues lie near $\operatorname{Re}\lambda=\beta$, and the continuous spectrum lies on the real axis.

\begin{figure}[htbp]
\centering
\begin{minipage}{0.47\textwidth}
    \centering
    \includegraphics[width=\linewidth]{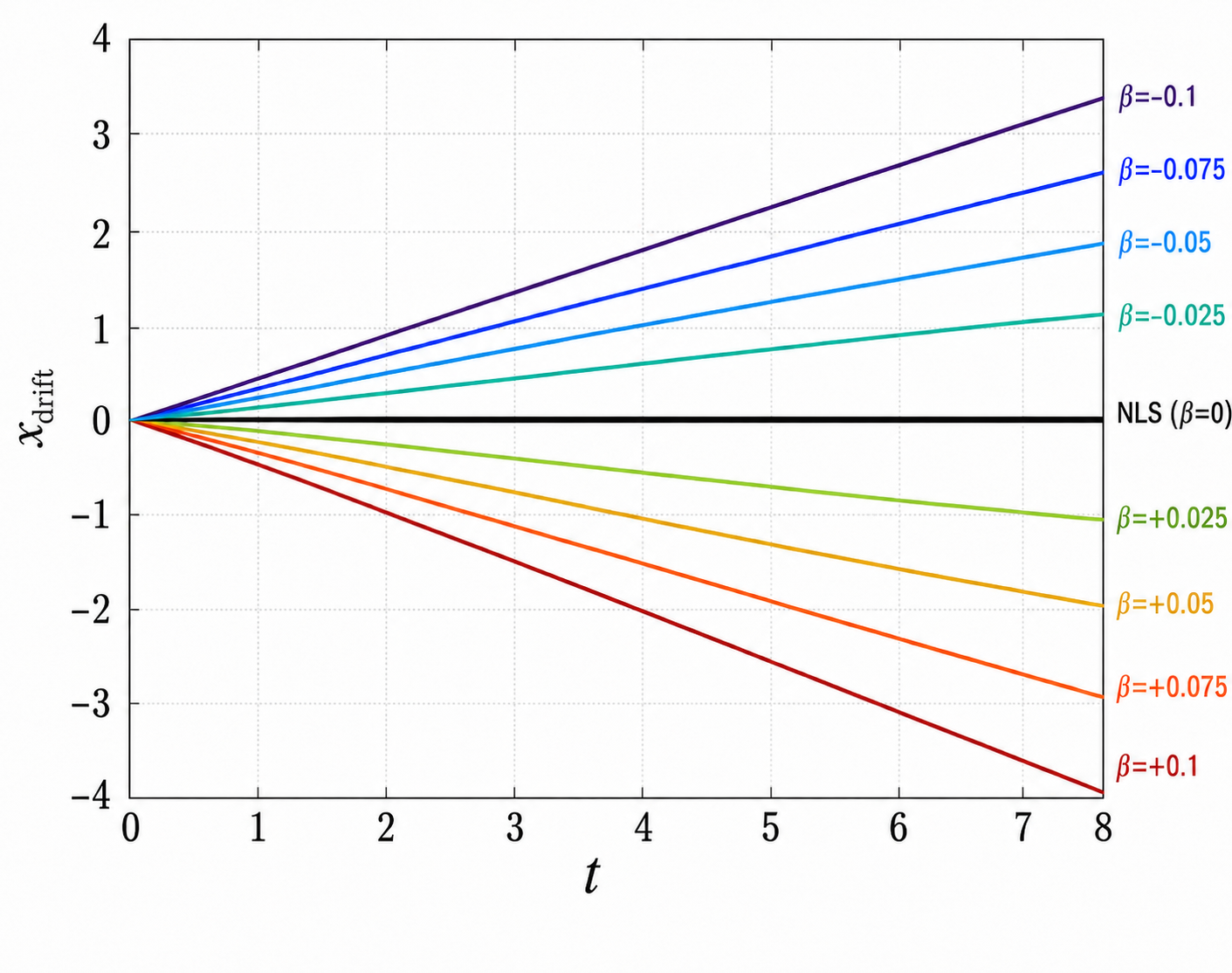}
    \caption*{(a)}
\end{minipage}
\begin{minipage}{0.43\textwidth}
    \centering
    \includegraphics[width=\linewidth]{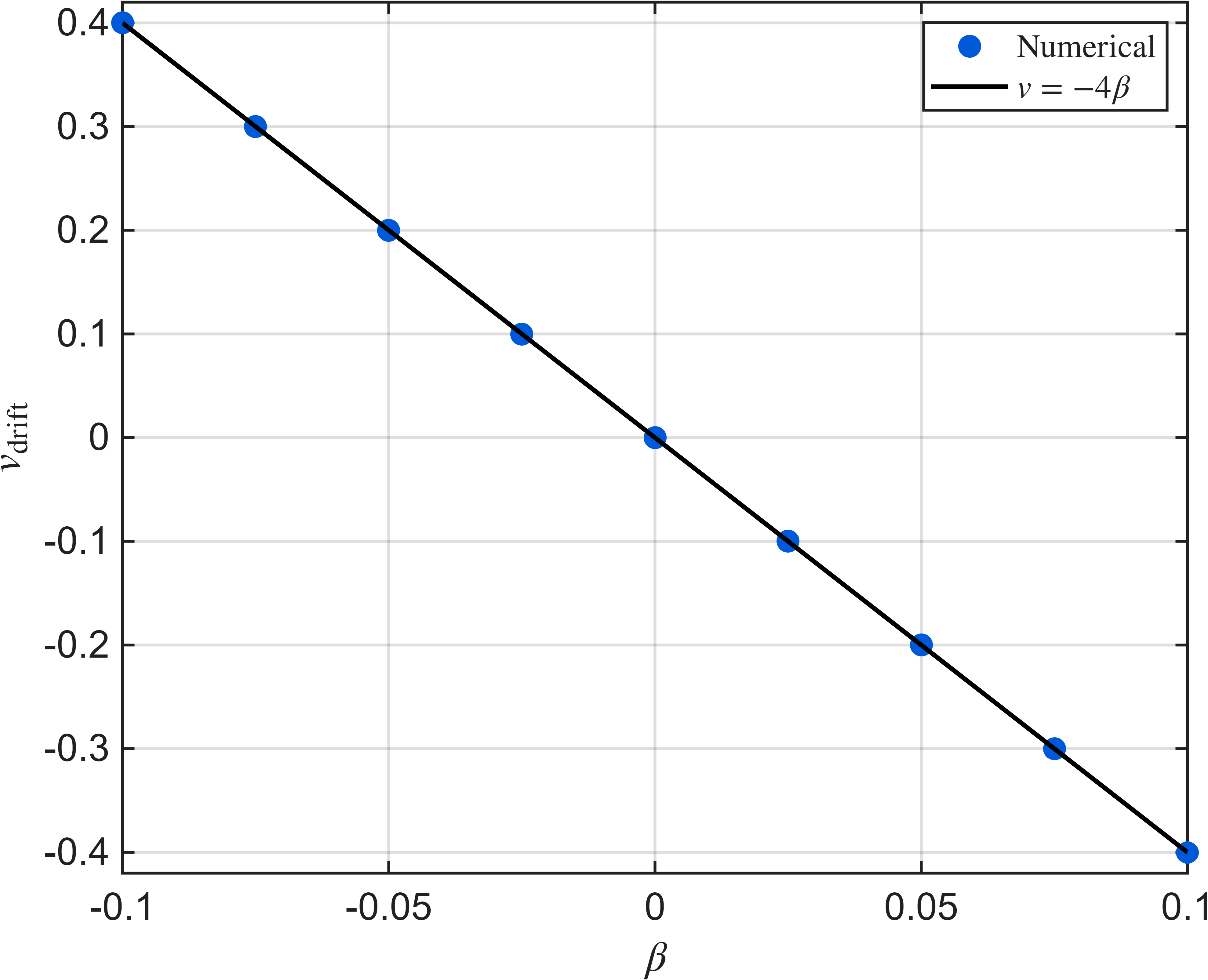}
    \caption*{(b)}
\end{minipage}
\caption{Numerical validation of the spatial drift induced by $\beta$. (a) Measured displacement for different values of $\beta$ compared with $x_{\mathrm{drift}}=-4\beta t$. (b) Measured drift velocity compared with $v_{\mathrm{drift}}=-4\beta$.}
\label{fig:beta_drift}
\end{figure}

We test the drift law~\eqref{eq:drift_law} for nine values of $\beta$ between $-0.10$ and $0.10$. We estimate the displacement from the cross-correlation of the KE and NLS intensity distributions and determine the velocity by a linear fit over $0.5\leq t\leq5$. The simulations use $N=4096$ spatial points and $\Delta t=0.001$. Figure~\ref{fig:beta_drift} compares the measured displacements and velocities with the predicted values. The numerical trajectories closely follow the predicted linear drift. A linear fit to the measured drift velocities gives
\[
v_{\mathrm{num}}=-4.000321\,\beta+2.82\times10^{-16}.
\]
The fit gives $R^2=0.99999998$, with a maximum absolute velocity error of $5.68\times10^{-5}$ relative to $v_{\mathrm{drift}}=-4\beta$.

To characterize the local spectral structure of individual intense events, we periodize selected portions of the wave field in the interaction region \cite{randoux2016inverse,wang2020numerical,tang2022local}. We evolve the KE equation on a domain of length $L=512$ with $N=4096$ grid points and $\Delta t=10^{-4}$, using Fourier pseudospectral discretization in space and the classical fourth-order Runge--Kutta (RK4) method in time. Following Refs.~\cite{chabchoub2011rogue,toenger2015emergent}, we classify a local peak as a rogue wave when
\[
A_{\mathrm{peak}}\geq2A_s,
\]
where $A_s$ is the mean of the highest one-third of the local envelope maxima at that time.

We select four events satisfying this criterion in $-12\leq x\leq12$. Following Ref.~\cite{randoux2016inverse}, we take the two side-lobe maxima surrounding each peak as the endpoints $x_L$ and $x_R$ of the extraction interval. The local period is $\Lambda=x_R-x_L$. Table~\ref{tab:local_intervals} lists the extraction intervals and periods.

\begin{table}[!htbp]
\centering
\caption{Selected intervals for the local Floquet analysis.}
\label{tab:local_intervals}

\renewcommand{\arraystretch}{1.25}
\begin{tabular*}{0.61\textwidth}
{@{\extracolsep{\fill}}ccc@{}}
\toprule
$t$ & Extraction interval $[x_L,x_R]$ & $\Lambda$ \\
\midrule
5.84  & $[-4.50,\;0.00]$    & 4.50  \\
10.86 & $[6.00,\;11.00]$    & 5.00  \\
12.63 & $[-8.00,\;-2.50]$   & 5.50  \\
14.33 & $[-6.75,\;0.875]$   & 7.625 \\
\bottomrule
\end{tabular*}
\end{table}

We periodically extend the field on each extraction interval and compute its Floquet spectrum using the Bloch condition \cite{deconinck2006computing}
\[
\phi(x+\Lambda)=e^{i\kappa\Lambda}\phi(x),
\]
where $\kappa$ is the Bloch wavenumber. An independent calculation based on explicit repetition of the extracted waveform \cite{randoux2016inverse} gives consistent Floquet band structures.

\begin{figure}[htbp]
    \centering
    \includegraphics[width=0.94\textwidth]
    {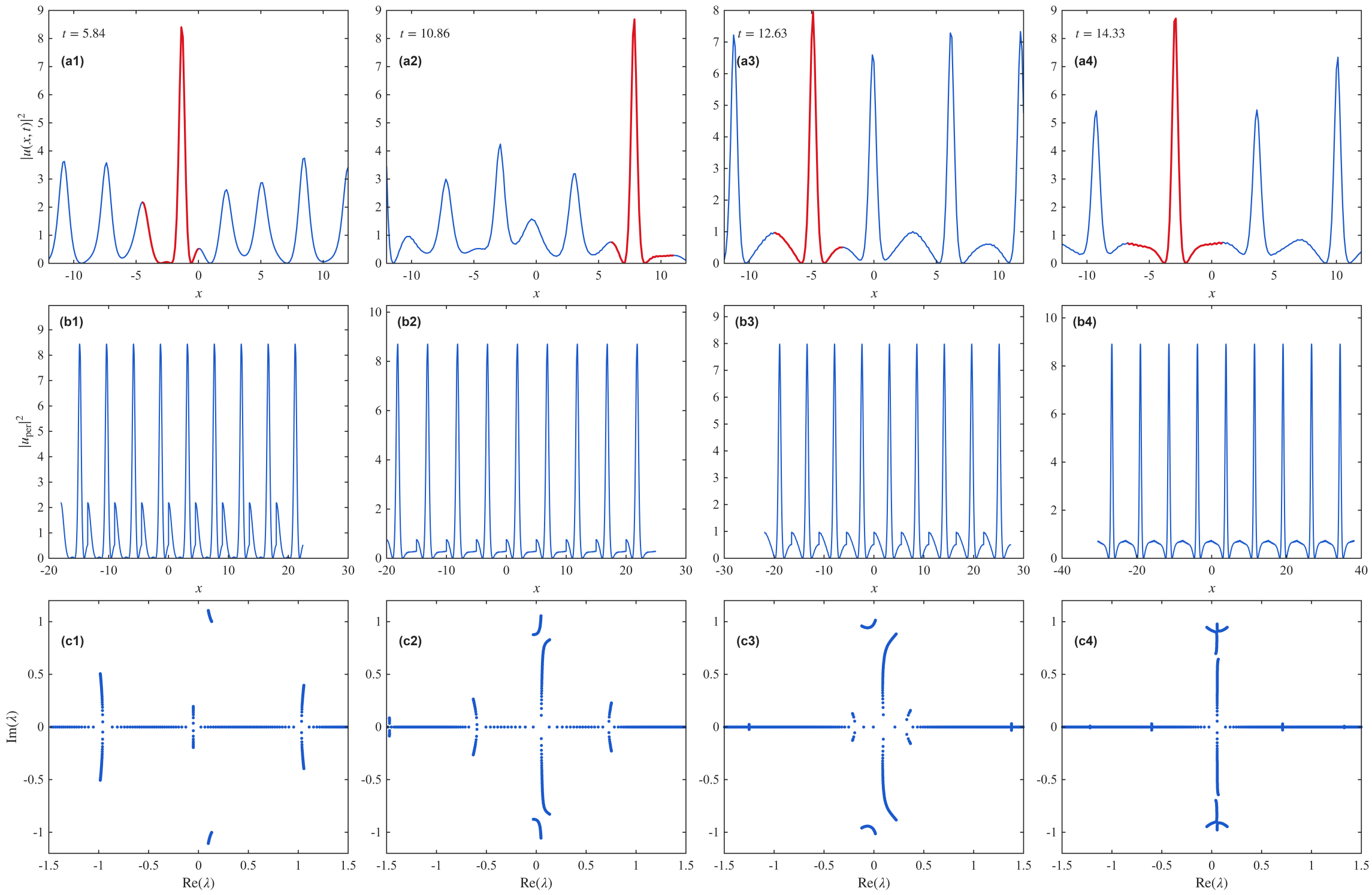}
    \caption{
    Local periodization and Floquet analysis of four selected rogue waves at $t=5.84$, $10.86$, $12.63$ and $14.33$. (a1)--(a4) Local intensity profiles $|u(x,t)|^2$ with the extracted intervals highlighted in red. (b1)--(b4) Periodic extensions of the extracted intervals. (c1)--(c4) Corresponding Floquet spectra.
    }
    \label{f11}
\end{figure}

The four local Floquet spectra in Figure~\ref{f11} differ in the positions and arrangement of their lateral branches. Among the selected events, the spectrum at $t=14.33$ is more compact and has less pronounced lateral branches. These spectra characterize different local waveforms; the global Lax spectrum of the full solution is invariant in time.

\begin{figure}[htbp]
    \centering
    \includegraphics[width=0.96\textwidth]
    {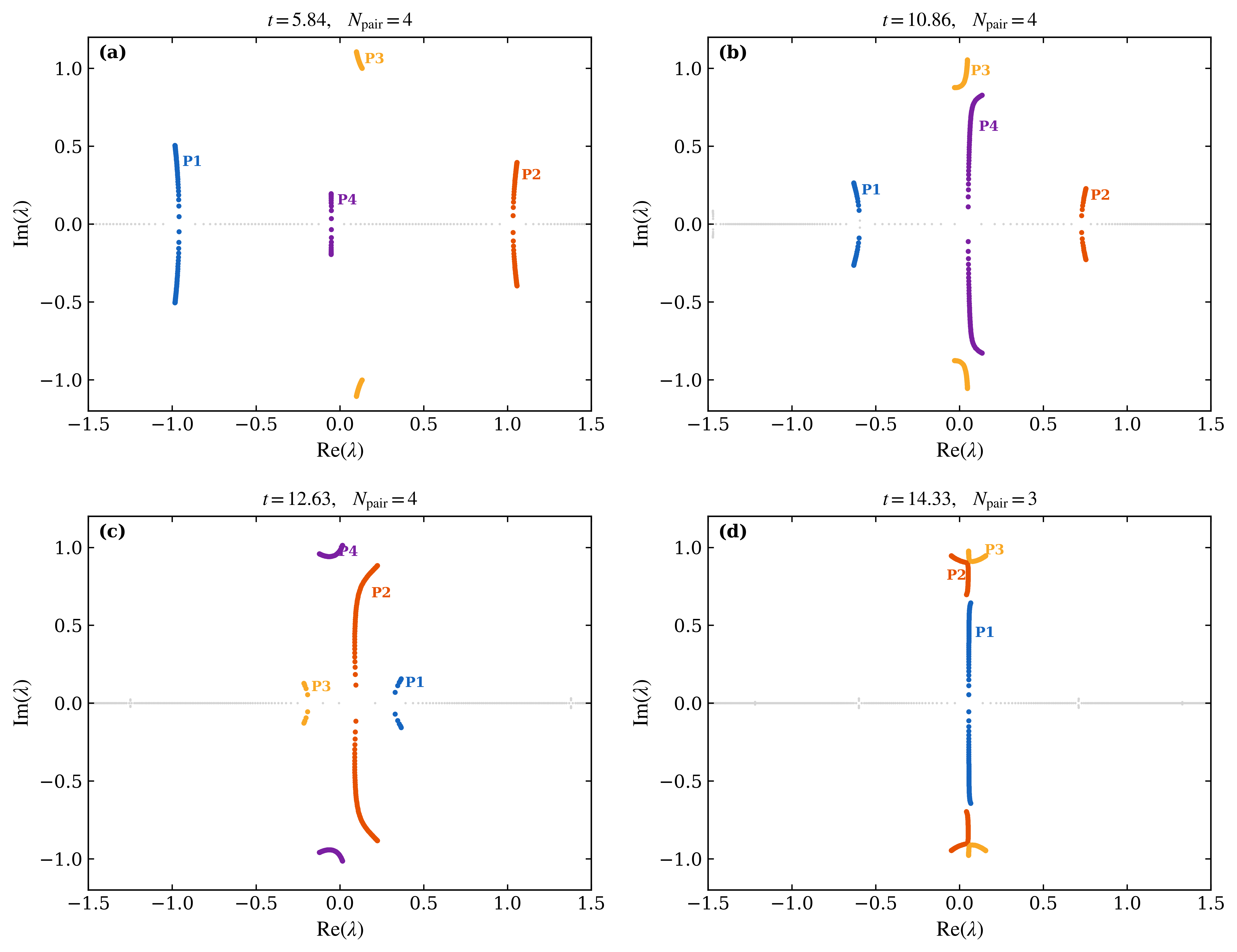}
    \caption{
    Automatic identification and conjugate pairing of the local Floquet branches. Branches shown in the same color are identified as one conjugate pair. The values of $N_{\mathrm{pair}}$ are $4$, $4$, $4$ and $3$ for $t=5.84$, $10.86$, $12.63$ and $14.33$, respectively.
    }
    \label{fig:auto_bands}
\end{figure}

We track the eigenvalues as the Bloch wavenumber $\kappa$ varies, split branches at discontinuous numerical jumps, and pair branches in the upper and lower half-planes by complex conjugation. For the reference periodization, this procedure gives
\[
N_{\mathrm{pair}}=4,\quad4,\quad4,\quad3
\]
for the four selected events. Here, $N_{\mathrm{pair}}$ is the number of identified conjugate band pairs and serves as a descriptor of the local Floquet spectrum. Figure~\ref{fig:auto_bands} shows the identified Floquet branches.

Following the sensitivity criterion in Ref.~\cite{randoux2016inverse}, we examine the effects of small changes in the extraction interval and period. We first shift the interval by $\pm0.05\Lambda$ at fixed length, then change its length to $0.95\Lambda$ and $1.05\Lambda$ at fixed center. For each modified spectrum, we compute nearest-neighbor distances to the reference spectrum in both directions. The 95th percentile of these distances measures the discrepancy between the two spectra. Table~\ref{tab:periodization_robustness} reports the larger discrepancy for the two signs of each perturbation. All reported values are below $0.10$.

\begin{table}[htbp]
\centering
\caption{Maximum errors in the Floquet spectra produced by $\pm5\%$ changes in the periodization parameters.}
\label{tab:periodization_robustness}
\begin{tabular}{ccc}
\toprule
Time & Window shift error & Period change error \\
\midrule
5.84  & 0.096 & 0.082 \\
10.86 & 0.079 & 0.088 \\
12.63 & 0.053 & 0.052 \\
14.33 & 0.007 & 0.011 \\
\bottomrule
\end{tabular}
\end{table}
    
\section{Random perturbations and ensemble statistics}\label{s4}

In this section, 2e compare dam-break evolution for uniform, Gaussian and Laplace perturbations with equal variance and the same correlation scale in the underlying Gaussian field.

\subsection{Random initial data}

The randomly perturbed box initial data is
\begin{equation*}
u(x,0)=
\begin{cases}
1+\epsilon\eta(x), & |x|\leq l,\\
0, & |x|>l,
\end{cases}
\end{equation*}
where $l=25$, $\eta(x)$ denotes the random perturbation and $\epsilon$ controls its strength. The values $\epsilon=0.05,\;0.10,\;0.15,\;0.20,\;0.30,\;0.40$ are considered. Three target marginal distributions are used:
\begin{equation}
\eta_U\sim U(-\sqrt{3},\sqrt{3}),\qquad \eta_G\sim\mathcal{N}(0,1),\qquad \eta_L\sim \mathrm{Laplace}\left(0,\frac{1}{\sqrt{2}}\right).\label{eq:noise_marginals}
\end{equation}
All three target distributions have zero mean and unit variance. The uniform distribution has bounded support, whereas the Gaussian and Laplace distributions have unbounded support. The Laplace tails decay more slowly than the Gaussian tails.

We generate a correlated Gaussian field $Z(x)$ by filtering Gaussian white noise in Fourier space. Its covariance is
\[
R_G(r)=\exp\left[-\left(\frac{r}{l_c}\right)^2\right],\qquad l_c=1,
\]
where $r$ denotes the spatial separation and $l_c$ is the latent correlation scale. The corresponding Fourier amplitude filter is
\[
H(k)=\exp\left[-\frac{(k l_c)^2}{8}\right],
\]
where $k$ is the spatial wavenumber. For background on generating correlated random fields, see Refs.~\cite{shinozuka1972digital,yamazaki1988digital}.

The latent Gaussian field is transformed to each target marginal according to
\[
\eta_j(x)=F_j^{-1}\!\left[\Phi(Z(x))\right],\qquad j=U,G,L,
\]
where $\Phi$ is the cumulative distribution function of the standard Gaussian distribution and $F_j^{-1}$ is the inverse cumulative distribution function of the target distribution \cite{yamazaki1988digital,grigoriu2007parametric}. The three perturbation fields thus share the same underlying Gaussian correlation scale.

\begin{figure}[htbp]
    \centering
    \includegraphics[width=\textwidth]{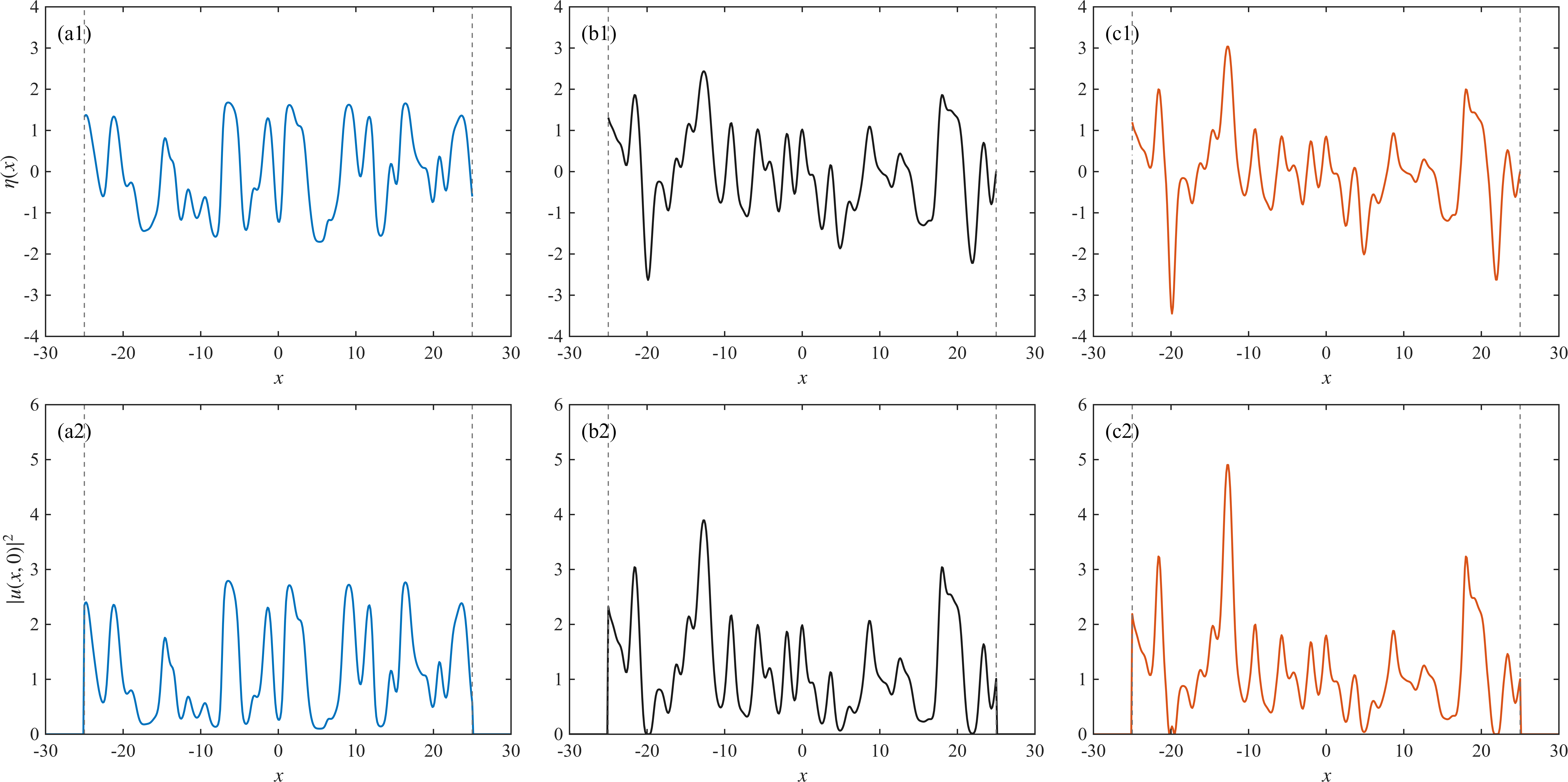}
    \caption{
    Representative random initial data at $\epsilon=0.40$. Panels (a1)--(c1) show the portions of $\eta(x)$ used in the box initial data for the uniform, Gaussian and Laplace distributions, respectively. Panels (a2)--(c2) show the corresponding initial intensities $|u(x,0)|^2$.
    }
    \label{fig:noise_initial}
\end{figure}

The target probability densities and tail probabilities are listed in \ref{app:noise_marginals}. Figure~\ref{fig:noise_initial} shows sample perturbations and the corresponding initial intensities for $\epsilon=0.40$. In these samples, the uniform perturbation has smaller local fluctuations, while the Gaussian and Laplace perturbations contain larger excursions.

\begin{figure}[htbp]
    \centering
    \includegraphics[width=\textwidth]{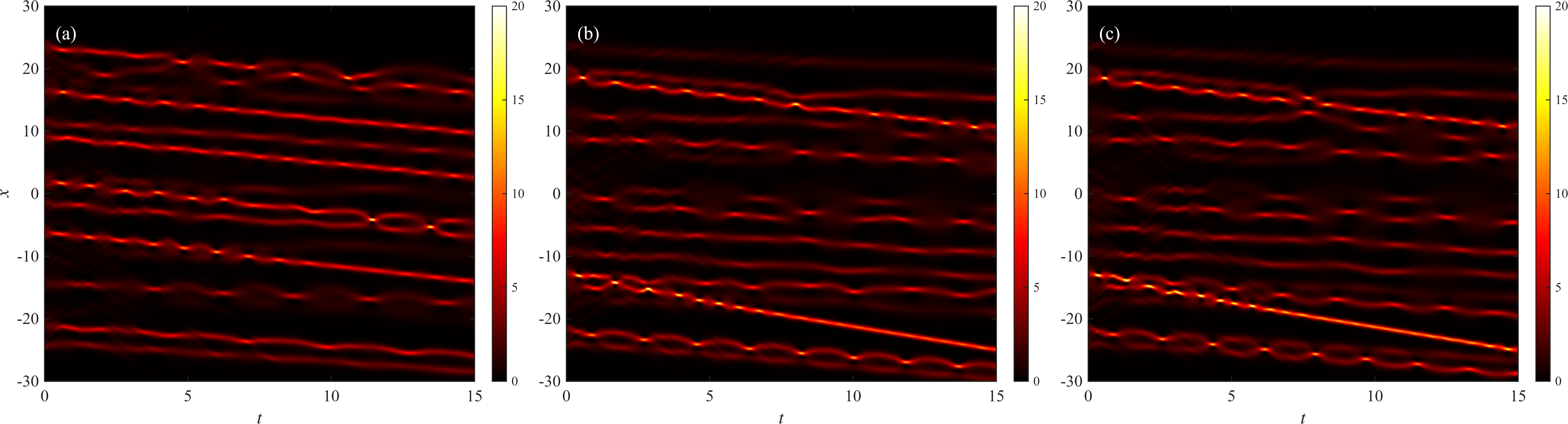}
    \caption{
    Representative evolutions of the intensity $|u(x,t)|^2$ at $\epsilon=0.40$ for (a) uniform, (b) Gaussian and (c) Laplace perturbations. The same scale is used in all panels.
    }
    \label{fig:noise_dynamics}
\end{figure}

For each distribution and each value of $\epsilon$, we generate 100 realizations and evolve them to $t_{\max}=15$ with $\beta=0.05$. The simulations use a domain of length $L=512$, $N=4096$ spatial grid points and $\Delta t=0.001$, with Fourier pseudospectral discretization in space and the classical fourth-order RK4 method in time. We apply the $2/3$ dealiasing rule to the nonlinear terms. The sample evolutions in Figure~\ref{fig:noise_dynamics} have similar overall dam-break patterns but differ in local peak intensities and interactions. We quantify these differences through ensemble statistics.

\subsection{Initial extremes and nonlinear amplification}

To distinguish large initial fluctuations from subsequent amplification, we compare the maximum intensity at $t=0$ with that reached during the evolution. We evaluate all spatial statistics in the moving observation window
\begin{equation*}
\mathcal{D}(t)=\left\{x:\left|x-x_{\mathrm{drift}}(t)\right|\leq40\right\},
\end{equation*}
where $x_{\mathrm{drift}}(t)=-4\beta t$ is the deterministic drift in Equation~\eqref{eq:drift_law}. This window excludes the nearly zero background far from the interaction region.

For each realization, the initial maximum intensity is
\[
I_{\max}^{(0)}=\max_{x\in\mathcal{D}(0)}|u(x,0)|^2.
\]
The maximum intensity attained during the simulated evolution is
\[
I_{\max}^{(\mathrm{dyn})}=\max_{\substack{t>0\\x\in\mathcal{D}(t)}}|u(x,t)|^2.
\]
Define the relative amplification factor
\[
G=\frac{I_{\max}^{(\mathrm{dyn})}}{I_{\max}^{(0)}}.
\]
Thus, $G$ compares the maximum intensity attained during the simulation with the initial maximum intensity. We also characterize intensity fluctuations through the normalized fourth moment (kurtosis)
\[
K(t)=\frac{\langle |u|^4\rangle}{\langle |u|^2\rangle^2},
\]
where $\langle\cdot\rangle$ denotes spatial averaging over $\mathcal{D}(t)$. The maximum value for each realization is
\[
K_{\max}=\max_{t>0}K(t).
\]

\begin{figure}[htbp]
    \centering
    \includegraphics[width=0.96\textwidth]
    {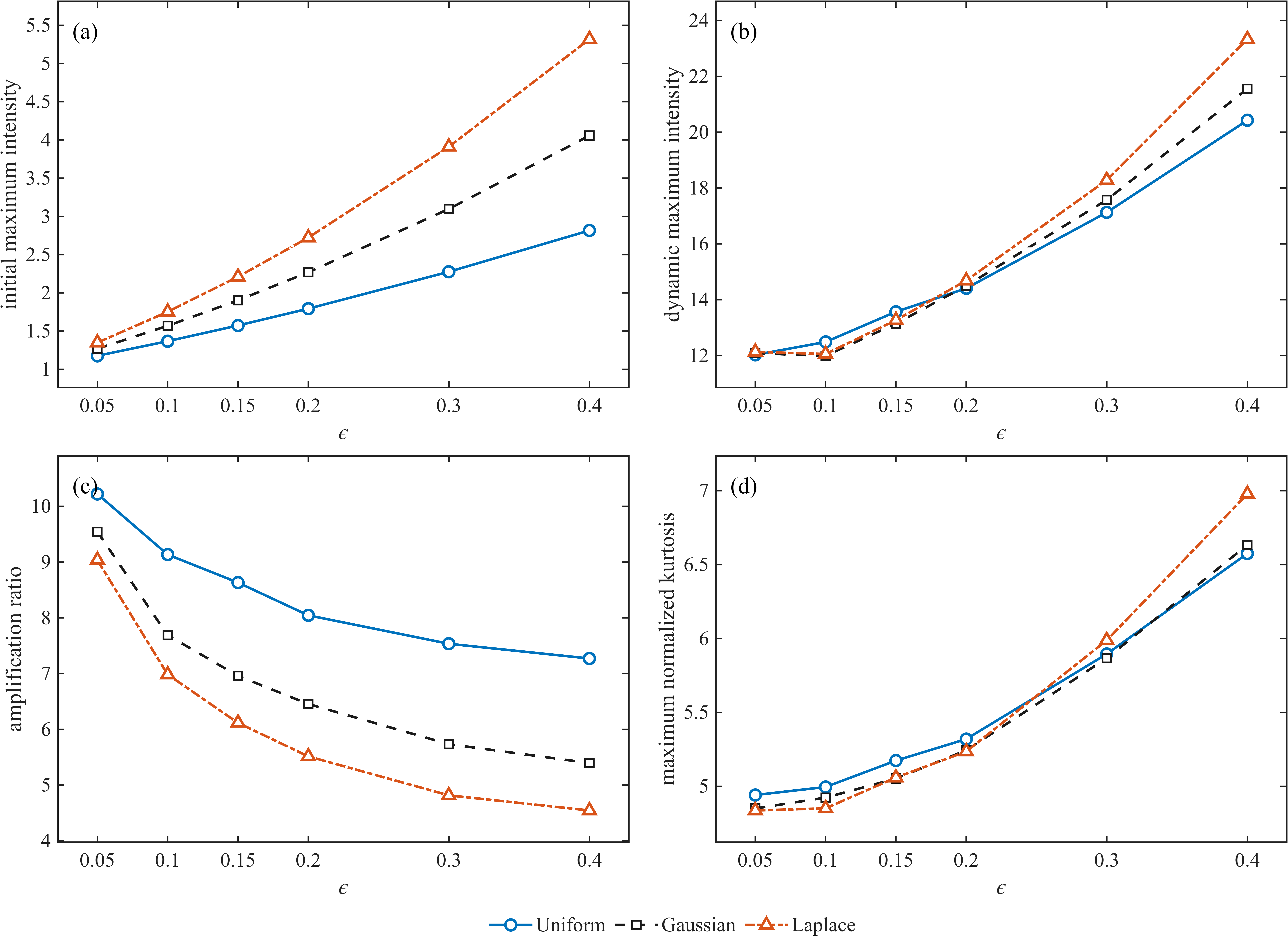}
    \caption{
    Ensemble statistics as functions of the perturbation strength $\epsilon$ for uniform, Gaussian and Laplace perturbations. (a) Mean initial maximum intensity $I_{\max}^{(0)}$. (b) Mean maximum intensity $I_{\max}^{(\mathrm{dyn})}$ attained during the evolution. (c) Mean relative amplification factor $G$. (d) Mean maximum normalized kurtosis $K_{\max}$. Each point represents the mean over 100 realizations.
    }
    \label{fig:ensemble_statistics}
\end{figure}

Figure~\ref{fig:ensemble_statistics} shows the ensemble statistics. The mean initial maximum intensity increases with $\epsilon$ for all three distributions and is largest for Laplace perturbations, followed by Gaussian and uniform perturbations. The differences increase with $\epsilon$.

The mean maximum intensities attained during the evolution are similar for weak and moderate perturbations. At the larger values of $\epsilon$ considered, the Laplace perturbations give the largest mean maxima. The ensemble mean of $G$ decreases as $\epsilon$ increases, with the largest decrease for Laplace perturbations. For moderate and strong perturbations, the mean factor is largest for the uniform distribution and smallest for the Laplace distribution. Thus, the larger maxima for Laplace perturbations coexist with smaller relative amplification, consistent with their larger initial extremes. The ensemble mean of $K_{\max}$ is similar for the three distributions at small $\epsilon$. The differences increase with $\epsilon$, and Laplace perturbations give the largest values at the stronger perturbation levels. This trend is consistent with more pronounced intensity intermittency.

\begin{figure}[htbp]
    \centering
    \includegraphics[width=0.96\textwidth]
    {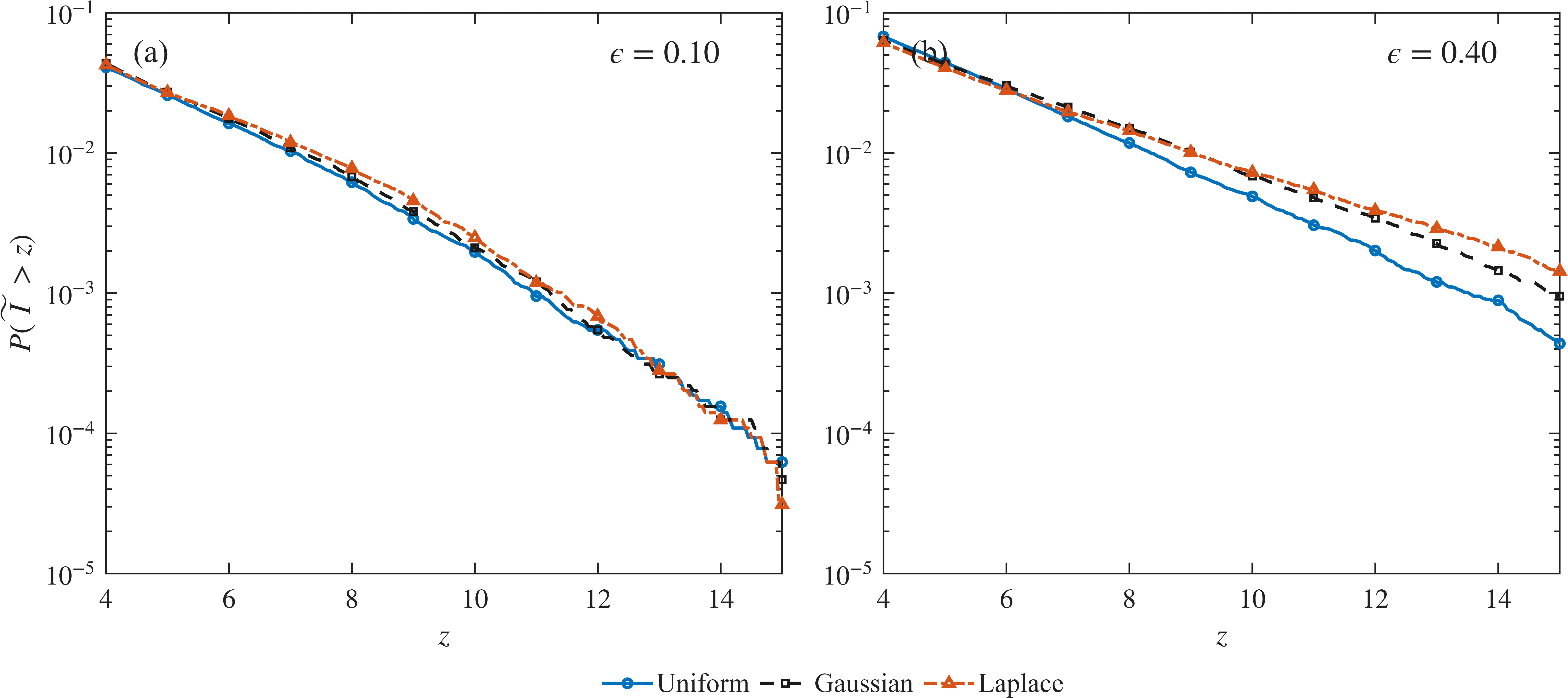}
    \caption{
    CCDFs of the normalized intensity $\widetilde{I}=|u|^2/\langle|u|^2\rangle$ at the final simulation time for (a) $\epsilon=0.10$ and (b) $\epsilon=0.40$. Results are based on 100 realizations for each perturbation distribution.
    }
    \label{fig:final_intensity_ccdf}
\end{figure}

At the final simulation time $t=15$, define the normalized intensity
\[
\widetilde{I}=\frac{|u|^2}{\langle |u|^2\rangle},
\]
where $\langle\cdot\rangle$ denotes spatial averaging over $\mathcal{D}(15)$. The complementary cumulative distribution function (CCDF), $P(\widetilde{I}>z)$, gives the exceedance probability at threshold $z$. Figure~\ref{fig:final_intensity_ccdf} shows the empirical CCDFs. At $\epsilon=0.10$, the empirical CCDFs nearly coincide. At $\epsilon=0.40$, the exceedance probability at large $z$ is highest for Laplace perturbations, followed by Gaussian and uniform perturbations. The dependence on the marginal distribution is therefore more pronounced at $\epsilon=0.40$ than at $\epsilon=0.10$.

\subsection{Lax spectra of randomly perturbed initial data}

For each realization, we compute the Lax spectrum of the initial data on a window of length $L_s=128$ using $N_s=1024$ spatial samples. For the unperturbed box, the discrete eigenvalues lie on the reference line $\operatorname{Re}\lambda=\beta$. The displacement of an eigenvalue from this line is defined as $d(\lambda) =\left|\operatorname{Re}\lambda-\beta\right|$. An Eigenvalue is regarded as displaced when $d(\lambda)>\delta_\lambda$, where $\delta_\lambda=0.005$ accounts for the numerical deviation from $\operatorname{Re}\lambda=\beta$ in the unperturbed box calculation. The retained eigenvalues satisfy
\[
0.05<\operatorname{Im}\lambda\leq2, \qquad |\operatorname{Re}\lambda-\beta|\leq1.5.
\]

We measure the fraction of retained spectral points satisfying $d(\lambda)>\delta_\lambda$ and the mean displacement of these points from the reference line. Both quantities are averaged over 100 realizations for each distribution and each value of $\epsilon$.

\begin{figure}[htbp]
    \centering
    \includegraphics[width=0.96\textwidth]
    {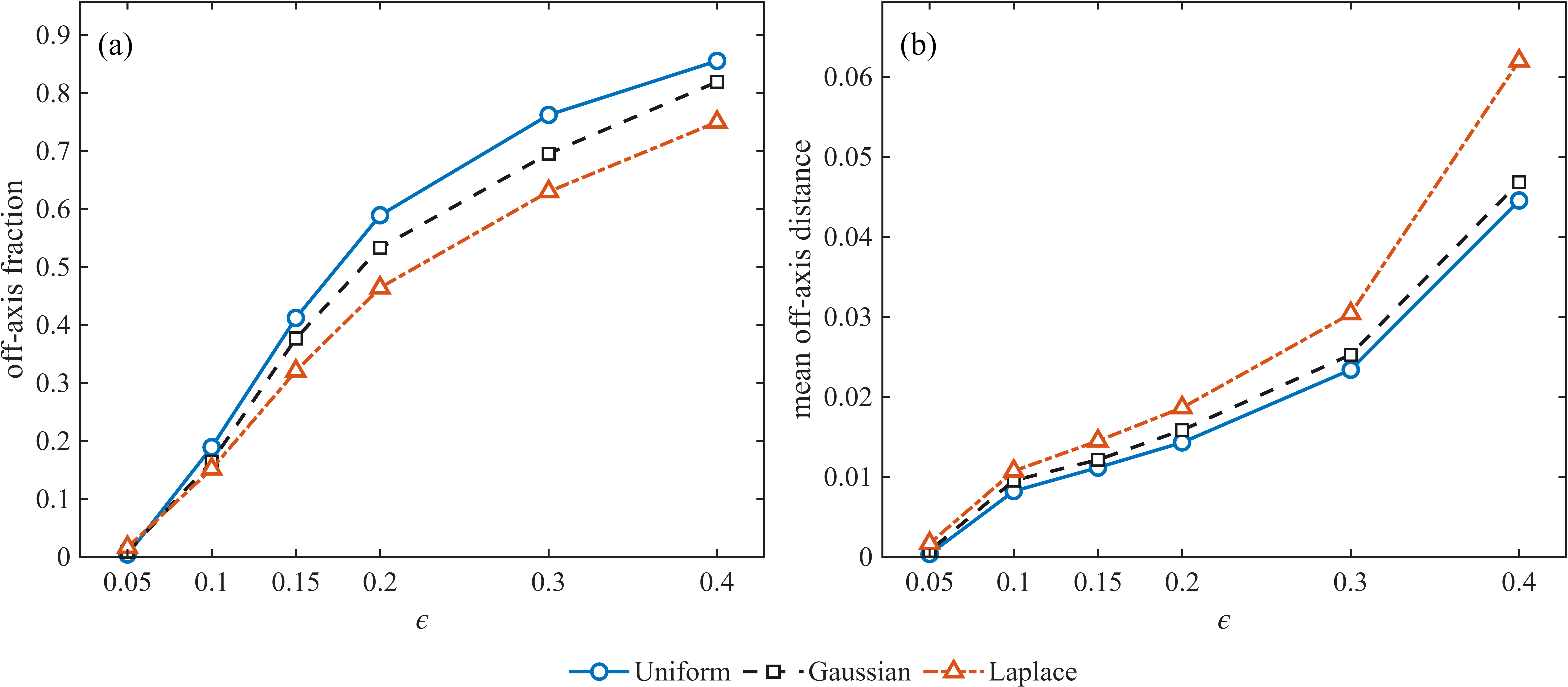}
    \caption{
    Ensemble statistics of the Lax spectrum versus perturbation strength. (a) Fraction of eigenvalues displaced from $\operatorname{Re}\lambda=\beta$ beyond the numerical tolerance. (b) Mean displacement of these points. Results are averaged over 100 realizations.
    }
    \label{fig:spectral_statistics}
\end{figure}

Figure~\ref{fig:spectral_statistics}(a) shows that the fraction of displaced spectral points increases with $\epsilon$ for all three distributions. This fraction is generally larger for uniform and Gaussian perturbations than for Laplace perturbations. The mean displacements in Figure~\ref{fig:spectral_statistics}(b) are similar at small $\epsilon$ and differ more at larger $\epsilon$. For strong perturbations, the mean displacement is largest for Laplace perturbations and smallest for uniform perturbations. The displaced fraction and the mean displacement therefore have different dependences on the perturbation distribution.

\section{Conclusions and discussion}\label{s5}

In this paper, the KE--NLS gauge transformation and Galilean invariance relate the two intensity evolutions by a spatial translation for the unperturbed real box. The MI growth rate, the number of discrete eigenvalues and their imaginary parts are independent of $\beta$. The parameter changes the phase, shifts the discrete spectrum along the real axis and determines the drift velocity. Numerical results agree with the predicted eigenvalue count, spectral shift and drift over the parameter ranges considered. The four selected rogue waves have distinct local Floquet spectra, while the global Lax spectrum is invariant in time. The local spectra show only small changes under the tested variations of the extraction interval and period.

For random perturbations with equal variance and the same underlying Gaussian correlation scale, the marginal distribution affects the maximum intensities and spectral displacements. At the stronger perturbation levels considered, Laplace perturbations give larger mean maximum intensities, higher final-time exceedance probabilities and larger mean spectral displacements, but smaller amplification relative to the initial maximum intensity. Larger absolute maxima thus need not indicate stronger relative amplification.

\section*{Funding}
This work is supported by the Kaiyuan International Mathematical Sciences Institute, the Fundamental Research Funds for the Central Universities (JKF2025077492962), and the Academic Excellence Foundation of BUAA for Ph.D. Students.

\section*{Declaration of competing interest}
The authors declare no competing interests.

\section*{Data availability}
The data that support the findings of this study are available from the corresponding author upon request.

\appendix
\renewcommand{\thesection}{Appendix~\Alph{section}}

\counterwithin{equation}{section}
\counterwithin{figure}{section}
\counterwithin{table}{section}

\renewcommand{\theequation}{\Alph{section}.\arabic{equation}}
\renewcommand{\thefigure}{\Alph{section}.\arabic{figure}}
\renewcommand{\thetable}{\Alph{section}.\arabic{table}}

\section{Scattering data and discrete eigenvalues} \label{app:analytical}

\subsection{Derivation of the scattering data for the box potential} \label{app:scattering_derivation}

The gauge transformation also relates the spatial spectral problems of the KE and focusing NLS equations. Define
\[
G(x,t)=
\begin{pmatrix}
e^{i\Theta/2} & 0\\
0 & e^{-i\Theta/2}
\end{pmatrix},
\qquad \boldsymbol{\psi}=G\boldsymbol{\phi}.
\]
Together with
\[
u=qe^{i\Theta},\qquad\Theta_x=2\beta|q|^2,
\]
this transformation reduces Equation~\eqref{eq:KE_spatial_Lax} to
\[
\begin{pmatrix}
-\partial_x & q\\
q^* & \partial_x
\end{pmatrix}
\boldsymbol{\phi}=i\lambda\boldsymbol{\phi}.
\]
This is the Zakharov--Shabat spectral problem for the focusing NLS equation.

Inside the box, the corresponding NLS initial data is
\[
q(x,0)=e^{-2i\beta(x+l)}.
\]
The phase is removed by the transformation
\[
\boldsymbol{\phi}=
\begin{pmatrix}
e^{-i\beta(x+l)} & 0\\
0 & e^{i\beta(x+l)}
\end{pmatrix}
\boldsymbol{\chi}.
\]
The transformed spectral problem is
\[
\begin{pmatrix}
-\partial_x & 1\\
1 & \partial_x
\end{pmatrix}
\boldsymbol{\chi}=i(\lambda-\beta)\boldsymbol{\chi}.
\]
The corresponding spectral parameter is therefore $\mu=\lambda-\beta$.

To derive the scattering data directly, we solve the KE spatial problem in the three regions
\[
x<-l,\qquad-l\leq x\leq l,\qquad x>l.
\]
For $x<-l$ and $x>l$, the potential satisfies $u(x,0)=0$. Equation~\eqref{eq:KE_spatial_Lax} then reduces to
\[
-\psi_{1,x}=i\lambda\psi_1,\qquad\psi_{2,x}=i\lambda\psi_2.
\]
For incidence from the left, the solutions in Regions I and III are
\[
\boldsymbol{\psi}^{\mathrm{I}}(x)=
\begin{pmatrix}
R(\lambda)e^{-i\lambda x}\\
e^{i\lambda x}
\end{pmatrix},
\qquad x<-l,
\]
and
\[
\boldsymbol{\psi}^{\mathrm{III}}(x)=
\begin{pmatrix}
0\\
T(\lambda)e^{i\lambda x}
\end{pmatrix},
\qquad x>l,
\]
where $R(\lambda)$ and $T(\lambda)$ denote the reflection and transmission coefficients, respectively. Inside the box, $u(x,0)=1$, so Equation~\eqref{eq:KE_spatial_Lax} becomes
\[
\begin{pmatrix}
-\partial_x+i\beta & 1\\
1 & \partial_x+i\beta
\end{pmatrix}
\boldsymbol{\psi}=i\lambda\boldsymbol{\psi}.
\]
For $\mu=\lambda-\beta$, the corresponding wavenumber is
\[
\kappa=\sqrt{1+\mu^2}=\sqrt{1+(\lambda-\beta)^2}.
\]
The general solution in Region II is
\[
\boldsymbol{\psi}^{\mathrm{II}}(x)=A
\begin{pmatrix}
1\\
i(\kappa+\mu)
\end{pmatrix}
e^{i\kappa x}+B
\begin{pmatrix}
1\\
-i(\kappa-\mu)
\end{pmatrix}
e^{-i\kappa x}.
\]
Continuity at $x=l$ gives
\[
A e^{i\kappa l}+B e^{-i\kappa l}=0,
\]
and
\[
iA(\kappa+\mu)e^{i\kappa l}-iB(\kappa-\mu)e^{-i\kappa l}=T(\lambda)e^{i\lambda l}.
\]
Continuity at $x=-l$ gives
\[
A e^{-i\kappa l}+B e^{i\kappa l}=R(\lambda)e^{i\lambda l},
\]
and
\[
iA(\kappa+\mu)e^{-i\kappa l}-iB(\kappa-\mu)e^{i\kappa l}=e^{-i\lambda l}.
\]

The four matching conditions give the reflection and transmission coefficients
\[
R(\lambda)=-\frac{\sin(2\kappa l)}{\kappa\cos(2\kappa l)-i\mu\sin(2\kappa l)}e^{-2i\lambda l},
\]
and
\[
T(\lambda)=\frac{\kappa}{\kappa\cos(2\kappa l)-i\mu\sin(2\kappa l)}e^{-2i\lambda l}.
\]
With the normalization used in the main text, the scattering coefficients satisfy
\[
a_{\mathrm{KE}}(\lambda)=\frac{1}{T(\lambda)},\qquad b_{\mathrm{KE}}(\lambda)=\frac{R(\lambda)}{T(\lambda)}.
\]
Therefore,
\[
a_{\mathrm{KE}}(\lambda)=\frac{\kappa\cos(2\kappa l)-i\mu\sin(2\kappa l)}{\kappa}e^{2i\lambda l},
\]
and
\[
b_{\mathrm{KE}}(\lambda)=-\frac{\sin(2\kappa l)}{\kappa}.
\]

For the real NLS box potential, the corresponding scattering coefficients are
\[
a_{\mathrm{NLS}}(\mu)=\frac{\kappa\cos(2\kappa l)-i\mu\sin(2\kappa l)}{\kappa}e^{2i\mu l},
\]
and
\[
b_{\mathrm{NLS}}(\mu)=-\frac{\sin(2\kappa l)}{\kappa}.
\]
Since $\lambda=\mu+\beta$, the KE and NLS scattering data satisfy
\[
a_{\mathrm{KE}}(\lambda)=e^{2i\beta l}a_{\mathrm{NLS}}(\lambda-\beta),
\]
and
\[
b_{\mathrm{KE}}(\lambda)=b_{\mathrm{NLS}}(\lambda-\beta).
\]
Thus, the KE spectrum is the real NLS box spectrum shifted by $\beta$. The factor $e^{2i\beta l}$ leaves the zeros of $a_{\mathrm{KE}}$ unchanged.

\subsection{Counting the discrete eigenvalues}\label{app:discrete_count}

To derive \eqref{eq:eigenvalue_count} from \eqref{eq:eigenvalue_condition}, set
\[
\nu=\sin\theta,\qquad 0<\theta<\frac{\pi}{2}.
\]
The discrete eigenvalue condition then becomes
\[
\cos\theta\cos(2l\cos\theta)+\sin\theta\sin(2l\cos\theta)=0,
\]
or equivalently,
\[
\cos\left(2l\cos\theta-\theta\right)=0.
\]
The roots satisfy
\[
2l\cos\theta-\theta=\left(m+\frac{1}{2}\right)\pi,\qquad m=0,1,2,\ldots.
\]
Define
\[
F(\theta)=2l\cos\theta-\theta.
\]
The function $F$ is strictly decreasing on $(0,\pi/2)$ since
\[
F'(\theta)=-2l\sin\theta-1<0.
\]
Its endpoint values are
\[
F(0)=2l,\qquad F\left(\frac{\pi}{2}\right)=-\frac{\pi}{2}.
\]
Each admissible value of $m$ therefore corresponds to at most one discrete eigenvalue. A root with $0<\nu<1$ exists when
\[
\left(m+\frac{1}{2}\right)\pi<2l.
\]
If equality holds, the corresponding root has $\nu=0$ and does not represent a discrete eigenvalue in the upper half-plane.

The number of discrete eigenvalues is consequently
\[
N(l)=
\begin{cases}
0,&0<l\leq\dfrac{\pi}{4},
\\[3mm]
n,&\dfrac{(2n-1)\pi}{4}<l\leq\dfrac{(2n+1)\pi}{4}, \qquad n=1,2,\ldots.
\end{cases}
\]
This is the counting formula~\eqref{eq:eigenvalue_count}.

\section{Numerical methods} \label{app:numerical_methods}

\subsection{Fourier collocation method}
\label{sec61}

We compute the KE Lax spectrum by Fourier collocation \cite{yang2010nonlinear}. The spatial problem~\eqref{eq:KE_spatial_Lax} is
\[
\begin{pmatrix}
-\partial_x+i\beta|u|^2 & u\\[1mm]
u^* & \partial_x+i\beta|u|^2
\end{pmatrix}
\boldsymbol{\psi}=i\lambda\boldsymbol{\psi}.
\]

We discretize this problem on the finite interval
\[
x\in\left[-\frac{L}{2},\frac{L}{2}\right]
\]
using $2N+1$ Fourier modes. The eigenfunctions and the potential are expanded as
\[
\psi_1=\sum_{n=-N}^{N}a_{1n}e^{ink_0x},\qquad\psi_2=\sum_{n=-N}^{N}a_{2n}e^{ink_0x},
\]
\[
u=\sum_{n=-N}^{N}b_ne^{ink_0x},\qquad u^*=\sum_{n=-N}^{N}b_n^*e^{-ink_0x},
\]
and
\[
|u|^2=\sum_{n=-N}^{N}c_ne^{ink_0x},
\]
where $k_0=\frac{2\pi}{L}$.

Substituting the Fourier expansions gives a matrix eigenvalue problem for the coefficients $\{a_{1n},a_{2n}\}$:
\[
\begin{pmatrix}
-B_1+i\beta B_2 & B_3\\
B_3^* & B_1+i\beta B_2
\end{pmatrix}
\begin{pmatrix}
A_1\\[1mm]
A_2
\end{pmatrix}
=i\lambda
\begin{pmatrix}
A_1\\[1mm]
A_2
\end{pmatrix},
\]
where
\[
A_1=(a_{1,-N},a_{1,-N+1},\ldots,a_{1,N})^T,
\]
\[
A_2=(a_{2,-N},a_{2,-N+1},\ldots,a_{2,N})^T,
\]
and
\[
B_1=ik_0\,\mathrm{diag}(-N,-N+1,\ldots,N-1,N).
\]

The convolution matrices are
\[
\setlength{\arraycolsep}{3pt}
B_2=
\scalebox{0.85}{$
\begin{pmatrix}
c_0 & c_{-1} & \cdots & c_{-N} &  &  &  \\
c_1 & c_0 & c_{-1} & \ddots & \ddots &  &  \\
\vdots & c_1 & c_0 & \ddots & \ddots & \ddots &  \\
c_N & \ddots & \ddots & \ddots & \ddots & c_{-N} &  \\
 & c_N & \ddots & \ddots & \ddots & \ddots & \vdots \\
 &  & \ddots & \ddots & \ddots & \ddots & c_{-1} \\
 &  &  & c_N & \cdots & c_1 & c_0
\end{pmatrix}
$},
\]
and
\[
B_3=
\scalebox{0.85}{$
\begin{pmatrix}
b_0 & b_{-1} & \cdots & b_{-N} &  &  &  \\
b_1 & b_0 & b_{-1} & \ddots & \ddots &  &  \\
\vdots & b_1 & b_0 & \ddots & \ddots & \ddots &  \\
b_N & \ddots & \ddots & \ddots & \ddots & b_{-N} &  \\
 & b_N & \ddots & \ddots & \ddots & \ddots & \vdots \\
 &  & \ddots & \ddots & \ddots & \ddots & b_{-1} \\
 &  &  & b_N & \cdots & b_1 & b_0
\end{pmatrix}.
$}
\]
We compute the matrix eigenvalues and eigenvectors using the QR algorithm \cite{parlett2000qr}.

\subsection{Fourier pseudospectral method}\label{sec62}

We evaluate spatial derivatives using the discrete Fourier transform (DFT) and advance the solution with an explicit time-stepping method \cite{yang2010nonlinear}.

A uniform grid is introduced on the domain $[a,b]$ of length $L=b-a$:
\[
x_j=a+j\Delta x,\qquad j=0,\ldots,N-1,\qquad \Delta x=\frac{L}{N}.
\]
Let $u_j^n\approx u(x_j,t_n)$ denote the numerical approximation. The DFT and its inverse are
\[
\widehat{u}_m^n=\mathcal{F}[u^n]_m=\sum_{j=0}^{N-1}u_j^n e^{-ik_mx_j},
\]
and
\[
u_j^n=\mathcal{F}^{-1}[\widehat{u}^n]_j=\frac{1}{N}\sum_{m=0}^{N-1}\widehat{u}_m^n e^{ik_mx_j}.
\]
The discrete wavenumbers are
\[
k_m=
\begin{cases}
\dfrac{2\pi m}{L},&0\leq m<\dfrac{N}{2},
\\[2mm]
\dfrac{2\pi(m-N)}{L},&\dfrac{N}{2}\leq m<N.
\end{cases}
\]

The spatial derivatives are evaluated as
\[
(\partial_{xx}u)_j^n=\mathcal{F}^{-1}\left[-k_m^2\widehat{u}_m^n\right]_j,
\]
and
\[
(\partial_x|u|^2)_j^n=\mathcal{F}^{-1}\left[ik_m\mathcal{F}(|u^n|^2)_m\right]_j.
\]
Equation~\eqref{e1} then gives the semi-discrete system
\[
\frac{du_j}{dt}=i\left[(\partial_{xx}u)_j+2|u_j|^2u_j+4\beta^2|u_j|^4u_j-4i\beta(\partial_x|u|^2)_ju_j\right],\qquad j=0,\ldots,N-1.
\]

Let $\mathbf{u}=(u_0,\ldots,u_{N-1})^T$ and let $F(\mathbf{u})$ denote the right-hand side of the semi-discrete system. The classical RK4 update from $t_n$ to $t_{n+1}=t_n+\Delta t$ is
\[
\begin{aligned}
k_1&=\Delta t\,F(\mathbf{u}^n),\\
k_2&=\Delta t\,F\left(\mathbf{u}^n+\frac{1}{2}k_1\right),\\
k_3&=\Delta t\,F\left(\mathbf{u}^n+\frac{1}{2}k_2\right),\\
k_4&=\Delta t\,F(\mathbf{u}^n+k_3),\\
\mathbf{u}^{n+1}&=\mathbf{u}^n+\frac{1}{6}(k_1+2k_2+2k_3+k_4).
\end{aligned}
\]
In the ensemble simulations, we apply the $2/3$ dealiasing rule to the nonlinear terms and use no post-step filter.

\section{Marginal distributions of the random perturbations}\label{app:noise_marginals}

\begin{figure}[htbp]
    \centering
    \includegraphics[width=0.9\textwidth]{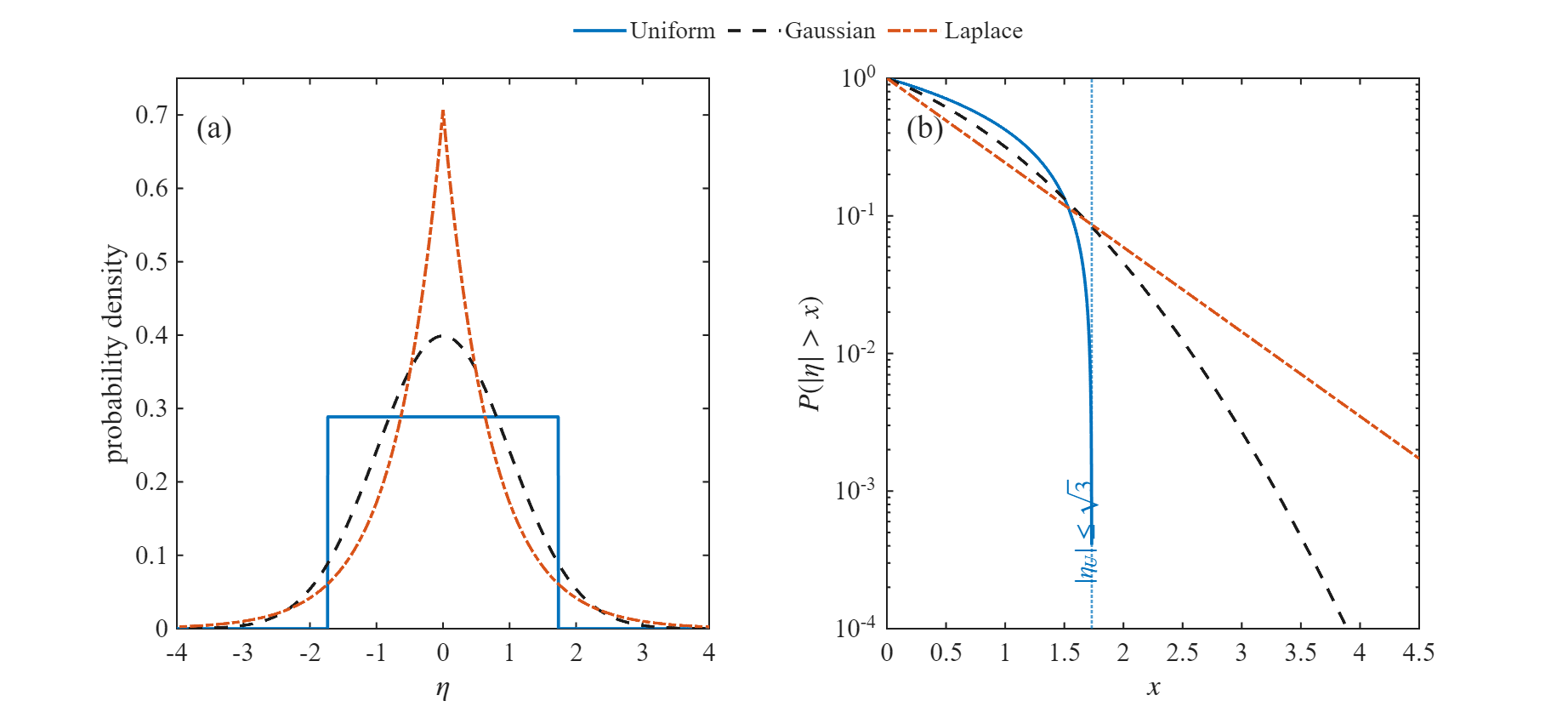}
    \caption{
    Comparison of the standardized target marginal distributions. (a) Probability density functions of the uniform, Gaussian and Laplace distributions. (b) Corresponding two-sided tail probabilities $\mathbb{P}(|\eta|>x)$ on a logarithmic scale. The vertical dotted line marks the uniform cutoff $x=\sqrt{3}$.
    }
    \label{fig:noise_marginal_tail}
\end{figure}

The uniform, Gaussian and Laplace target marginals used in Section~\ref{s4} have zero mean and unit variance before finite-sample recentering and rescaling. Their probability density functions are
\[
p_U(\eta)=
\begin{cases}
\dfrac{1}{2\sqrt{3}},&|\eta|\leq\sqrt{3},
\\[2mm]
0,&|\eta|>\sqrt{3},
\end{cases}
\]
\[
p_G(\eta)=\frac{1}{\sqrt{2\pi}}\exp\left(-\frac{\eta^2}{2}\right),
\]
and
\[
p_L(\eta)=\frac{1}{\sqrt{2}}\exp\left(-\sqrt{2}\,|\eta|\right).
\]
For $x\geq0$, the corresponding two-sided tail probabilities are
\[
\mathbb{P}(|\eta_U|>x)=
\begin{cases}
1-\dfrac{x}{\sqrt{3}},&0\leq x\leq\sqrt{3},
\\[2mm]
0,&x>\sqrt{3},
\end{cases}
\]
\[
\mathbb{P}(|\eta_G|>x)=\operatorname{erfc}\left(\frac{x}{\sqrt{2}}\right),
\]
and
\[
\mathbb{P}(|\eta_L|>x)=\exp\left(-\sqrt{2}\,x\right).
\]
Figure~\ref{fig:noise_marginal_tail} compares the tails of the three target distributions. At equal variance, the uniform distribution has bounded support, while the Laplace tails decay more slowly than the Gaussian tails.

\end{sloppypar}
\end{document}